\documentclass{aa}  

\usepackage{xcolor}
\usepackage{graphicx}
\usepackage{txfonts}
\usepackage{ulem}
\def\msun{M$_\odot$}

\newcommand{\iso}[1]{$^{#1}$}

\usepackage[]{hyperref}
\begin{document}

   \title{An AGB star as the source of the abundance pattern of hyper-metal-poor star HE~1327-2326}

   \titlerunning{AGB pollution for HE~1327-2326}


   \author{P. Gil-Pons
          \inst{1,4}
          \and
          S. W. Campbell\inst{2,3}
          \and
          C. L. Doherty\inst{2,3}
          \and
          M. Lugaro\inst{5,6,2}
          }

   \institute{ 
       EETAC, Universitat Politècnica de Catalunya, CBL, 08840 Castelldefels, Spain, 
    \email{pilar.gil@upc.edu}
    \and School of Physics and Astronomy, Monash University, Victoria 3800, Australia
    \and ARC Centre of Excellence for All Sky Astrophysics in Three Dimensions (ASTRO-3D), Melbourne, Australia
    \and Institut d'Estudis Espacials de Catalunya IEEC, Barcelona, Spain
    \and Konkoly Observatory, Research Centre for Astronomy and Earth Sciences, Konkoly Thege Miklós út 15-17, H-1121 Budapest, Hungary
    \and ELTE Eötvös Loránd University, Institute of Physics, Budapest 1117, Pázmány Péter sétány 1/A, Hungary
             }


 
  \abstract
    {Understanding the most metal-poor
   objects is a key to interpreting the nature of the first stars. 
   HE~1327-2326 (HE~1327), with metallicity [Fe/H]~$= -5.2$, is one of the most metal-poor stars detected and a candidate to be the offspring of the first stars. Numerous efforts have been made to match its abundance pattern, especially with high-mass stars undergoing supernova (SN) explosions.
   However, no model satisfactorily explains its entire surface chemical composition.}
   {The characteristic high CNO pattern with [N/Fe]~$>$~[C/Fe]~$>$~[O/Fe], the light element `slide' (between Na and Si), and the presence of Sr and Ba in HE~1327 is reminiscent of those asymptotic giant branch (AGB) stars that undergo third dredge-up, hot bottom burning, and s-processing -- suggesting that these stars may have been the source of the chemistry of the star. We aim to test this hypothesis.}
    {We assume that, where HE~1327 formed, the interstellar medium was well-mixed, and adopt an initial stellar composition based on the observed chemical evolution of the early universe. Zinc, which is enhanced in HE~1327, is well matched by this initial composition, as are the $\alpha$-elements. We calculated models of hyper-metal-poor AGB stars and compared the predicted chemical yields to the observed chemical pattern of HE~1327.}
   {We find our 3~\msun~ models match 13 of the 14 measured elements in HE1327, more than any model thus far. They are also consistent with the seven elements with upper limits. The only discrepancy is oxygen, underproduced by $0.5 - 1.0$~dex. For elements up to Zn, the match is comparable to that of the best-fitting, finely tuned SN models. Unlike the SN models, the AGB models also match Sr and Ba. We stress that the AGB scenario only requires standard stellar evolution without invoking exotic scenarios. Our model predicts high abundances of P and Pb, thus observations of these elements would be useful in testing the AGB scenario.}
  {We propose that HE~1327 is the oldest known object that shows nucleosynthetic evidence of the first AGB stars.
  With lifetimes as short as 200~Myr, these stars may have formed and polluted the universe very early. Recent Pop III star formation simulations support their formation, and their strong nitrogen production is qualitatively consistent with recent JWST observations showing high N/O ratios just 440~Myr after the Big Bang. Importantly, our results also suggest that the interstellar medium may have shown some degree of homogeneity and mixing even at these early epochs.}
  
\keywords{
    nuclear reactions, nucleosynthesis, abundances -- stars: evolution -- stars: Population II -- stars: AGB and post-AGB -- stars: carbon stars -- stars: individual: HE~1327-2326.}

   \maketitle
%

\section{Introduction}\label{sec:intro}

HE~1327-2326, with metallicity [Fe/H] = -5.2  \citep{ezzeddine2018}, is one of the most metal-poor stars currently known. Together with the other 13 stars detected with [Fe/H]~$<-4.5$, it belongs to a very early stellar generation. Interpreting the origin of these stars is crucial to understanding the very early universe.

At these extremely low metallicities, stars are almost exclusively carbon-enhanced (CEMP) but generally show no enhancements in elements heavier than iron, and are thus known as CEMP-no stars (e.g. \citealt{Yoon2016}). The current generalised conception is that CEMP stars with s-element enhancements were not able to form below this metallicity \citep{molaro2023} because it is expected that, in the very early universe, low- and intermediate-mass stars did not have enough time to form, evolve, and pollute other stars with s-process elements. Thus, massive stars are thought to have been the main contributors to CEMP-no abundance patterns. At these lowest metallicities it is also often assumed that the surface abundances of CEMP-no stars reflect the pollution from only a few -- or possibly only one -- progenitors. In the latter case, a single Pop III supernova yield is expected to explain the entire abundance pattern of a hyper-metal-poor star. We refer to this as the Pop~III supernova scenario. 

In the following, we first summarise the observed properties of HE~1327, as well as the various models which have attempted to explain the star's abundance pattern. We then describe our alternative scenario, which involves a hyper-metal-poor intermediate-mass ($\sim 3$~\msun{}) AGB star.

Since the discovery of HE~1327-2326 (\citealt{frebel2005}; hereafter HE~1327), the precise determination of its structural properties and surface abundances inspired great interest (e.g. \citealt{aoki2006}, \citealt{collet2006}, \citealt{frebel2006,frebel2008}, \citealt{bonifacio2012}, \citealt{yong13}, \citealt{ezzeddine2018}, \citealt{ezzeddine2019}, \citealt{molaro2023}).
Despite being a subgiant \citep{korn2009,mashonkina2017,brown2018,ezzeddine2018}, a study of its internal mixing suggested that the surface abundances of HE~1327 were only mildly depleted ($\simeq$0.2 dex; \citealt{korn2009}). Therefore, its surface abundances reflect its initial composition and thus it remains a most valuable relic of primitive stellar nucleosynthesis.

With a metallicity of [Fe/H] = -5.2 and a carbon abundance of [C/Fe$] = +3.49$~dex \citep{ezzeddine2019} HE~1327 is a CEMP star. Interestingly, it has consistently been classified as CEMP-no in the literature. This is despite it being clearly s-enhanced, with [Sr/Fe$] = +1.1$~dex \citep{frebel2005,frebel2008,aoki2006,ezzeddine2019}. The CEMP-no categorisation is likely due to the fact that the standard definition of CEMP-s stars requires Ba and Eu to be known \citep{beers2005}. Barium has only recently been measured in HE~1327 \citep{molaro2023}, and Eu still has not been detected. The categorisation is also motivated by HE~1327 being hyper metal-poor, like the majority of CEMP-no stars, in combination with it showing no evidence of binarity thus far (as opposed to most CEMP-s stars; \citealt{luc05, Hansen2016b}). 

The classification of HE~1327 as a CEMP-no, and the prevailing hypothesis that massive stars dominated the primitive IMF favoured its interpretation as the offspring of a single massive Pop. III star (e.g., \citealt{frebel2005}, \citealt{iwamoto2005}, \citealt{mey06},
\citealt{tominaga2007,tominaga2014}, \citealt{ezzeddine2019} and \citealt{jeena2023}). 
Zinc was only recently measured \citep{ezzeddine2018} and was fit by \cite{ezzeddine2019} using an aspherical Pop III supernova model (excluding Sr). 

The recent measurement of barium by \cite{molaro2023} showed that this s-process element is enhanced to a similar degree as Sr, with [Ba/Fe$]=+1.3$~dex. This suggests a reopening of the debate on the classification of HE~1327. Given the abundances and discussion above, and adding that nitrogen is highly enhanced, with [N/Fe$]=+3.98$~dex (and [N/C$]=+0.5$), we propose the classification of HE~1327 as a N-enhanced CEMP-s, or N-CEMP-s star (\citealt{izzard2009,pols2012}). We note that Eu should be measured to fit with the classic definition of (N)CEMP-s stars. Most importantly, it is crucial to reconsider the origin of HE~1327. 

As mentioned, the current dominant theory for the chemical pattern of HE~1327 is that it is wholly explained the descendant of a massive Pop III star. 
For example \citealt{iwamoto2005} proposed that HE~1327 was the offspring of a 25~\msun{} faint ($E<10^{51}\:erg$) core-collapse SN with mixing and fallback. 
Their model reproduces the Na, Mg, Al `slide' (where each heavier element has progressively lower abundance), and the Ca abundance, but fails to explain the [N/Fe]>[C/Fe]>[O/Fe] pattern and the formation of Ni, Zn, Sr and Ba. \citealt{takahashi2014} fitted HE~1327 with the yields from a low-energy rotating SN of 20 \msun{}. Their model reproduced C, O and the light-element slide but significantly underproduced N and Si (and did not attempt to match elements heavier than this).

\citealt{ezzeddine2019} considered an energetic aspherical bipolar jet SN. This model allows the fallback of matter into the emerging black hole and the release of ejecta along jets. The model yields are enriched in ashes up to Si-burning and very little Fe. Their best fit considers a 25~\msun{} model with artificially modified densities.
Their model succeeded in matching all observed elements except Sr (which was not reported). \citealt{ezzeddine2019} referred to \citealt{maeder2015} (who considered primordial models) and \citealt{choplin2017} (who considered models of [Fe/H]~$=-1.8$) to suggest that Sr could form in fast-rotating Pop III supernova progenitors. 
The CNO pattern, where N is highest in HE 1327, was not matched by the model (although we suggest N is within the model and observational uncertainties).

\citealt{jeena2023} proposed a rapidly rotating 12~\msun{} star, undergoing a quasi-chemically homogeneous state and diluting the wind and explosion ejecta with specific factors. The explosive yields of their models were computed assuming mixing and fallback as in \cite{maeda2003,umeda2003,tominaga2007} and \cite{ishigaki2014}. 
Similarly to the model of \citealt{ezzeddine2019}, their best-fitting model matched all of the elements except Sr (not reported). It shares the same issue as the other massive star/SN models, in that the N > C > O pattern is not reproduced. In addition, Zn is only marginally reproduced.
Finally, \cite{choplin2017} proposed massive rotating models to reproduce the heavy elements in HE~1327. However, those models underproduce N (compared to C and O), and, to our knowledge, no fit has been reported. 

In summary, for the Pop~III SN scenario, some of the fits have been quite successful, although they systematically disregarded the detection of [Sr/Fe] $\approx +1$~dex, and have not been able to match the CNO pattern reliably.

Apart from the primordial SN scenario, it has also been suggested that low-mass stars may have produced the abundance pattern of HE~1327. In this scenario, it is assumed that the low-mass star formed from an already-polluted gas cloud that had reached an [Fe/H$]=-5.2$. The pollution of the natal gas cloud is often assumed to have been contributed by a Pop~III SN, mixed with Big Bang material (e.g. \citealt{campbell2008}). The low-mass star then has to provide the enrichment of the elements above that initial composition. 

In particular, low-mass stars undergoing proton ingestion episodes (PIEs) have been proposed in this context, since they can provide Sr as well as enhanced CNO elements.
This low-mass scenario has been less successful than the SN scenario, however. For example, the 1~\msun{}, [Fe/H]~$=-6.5$ model in \cite{campbell2010}  fitted the observed abundances of HE~1327 within a factor of 4, but could not reproduce the [N/Fe] > [C/Fe] > [O/Fe] pattern and overproduced Sr and Ba by a large margin. \cite{cruz13} also attempted a match with a 1~\msun{}, $Z=10^{-8}$ model. That model could reproduce [Sr/Fe], but only at the expense of overproducing C, N and O by more than two orders of magnitude, and failing to match Na, Mg and Al, as well as Ca, Ti and Ni. \cite{cruz13} attributed the disagreements to the problems of PIE calculations and did not discard low-mass stars as candidates to interpret HE~1327.

For reference, in Section~\ref{sec:compareobs}, we quantify the fits to observations for a representative sample of literature models for both scenarios described above.

In the current study, we stress that the high C, N and O enrichment, the characteristic [N/Fe]~$>$~[C/Fe]~$>$~[O/Fe], the light-element (Na-Si) `slide', and the presence of s-process elements hint at an origin associated with the AGB phase of intermediate-mass stars, as suggested by e.g. \cite{campbell_phd,cruz13,gilpons2022}. As with the PIE low-mass scenario, this scenario requires the natal cloud to be already polluted, since low- and intermediate-mass stars do not produce iron (for example). 

At variance with the PIE scenario, we do not assume an inhomogeneous interstellar medium (ISM) -- we make the assumption that at the time HE~1327 formed, the ISM had undergone some chemical evolution, at least locally in the formation cloud of HE~1327. By that time, core-collapse supernovae had already enriched the ISM with $\alpha-$elements and Fe. 

Thus we propose that HE~1327, an ancient low-mass star ($\sim 0.8$ \msun), received its unusual abundance pattern through the superposition of the yield of an AGB star of intermediate mass ($\sim 3$~\msun{}) onto the prevailing abundance pattern of the interstellar medium at the time (at [Fe/H]~$ \sim -5.2$~dex). This could have happened either through an intermediate-mass star polluting the natal gas cloud of HE~1327, or through mass transfer from an intermediate-mass binary companion in a wide orbit (binarity in HE~1327 is neither confirmed nor ruled out). Either way, an intermediate-mass star provides all the abundance enhancements (CNO, the light element slide, and s-process) above the background natal composition.

To test this hypothesis, we (i) estimate the prevailing abundance pattern at the time HE~1327 formed, (ii) calculate the chemical yield of such an intermediate-mass star that formed from this material (Section~\ref{sec:models}), (iii) assume some degree of dilution in the ISM where HE~1327 formed, and (iv) compare the resultant theoretical composition to the observed chemical pattern (Section~\ref{sec:compareobs}). 

\section{Intermediate-mass hyper metal-poor stellar models}\label{sec:models}

\subsection{Stellar structure code}

For the structural evolution calculations, we use the Monash/Mount Stromlo stellar code \textsc{monstar} (e.g., \citealt{lat86,karakas2007,campbell2008,doherty2010,constantino2014}), specifically the version described in \citealt{gilpons2022}. This code considers the nuclei required to calculate the structural evolution. The input physics that are particularly relevant to the current work are as follows. Convection is treated via the mixing-length theory \citep{bom58} with mixing-length parameter $\alpha=1.75$. Convective boundaries are determined by the Schwarzschild criterion modified by the search for convective neutrality approach described in \citealt{frost1996}, with the option to use diffusive overshoot \citep{Herwig1997, Constantino2015}. Composition-dependent low-temperature opacities were used, with tables from \textsc{aesopus} \citep{lederer2009,marigo2009,constantino2014}.
Mass-loss rates at very low metallicities are highly uncertain because of the lack of observational calibration. We adopted the \citealt{bloecker1995} mass-loss formalism with $\eta=0.05$, as values between $\sim 0.01$ and 0.1 agree with observations at the solar and moderately metal-poor regimes and are frequently reported in the literature of very and extremely metal-poor stellar models (e.g. \citealt{ventura2001, her04b, ritter12, gilpons2022}). 

Models were calculated for masses between 3.0 and 3.5~\msun, with and without convective overshooting ($f_{OV}=0.01$). When overshooting was used, it affected all the convective boundaries. Table \ref{tab:evol1} summarises some key parameters and results from our models.

\begingroup
\setlength{\tabcolsep}{4pt}
\begin{table}[h]
   \begin{center}
         \label{tab:mod}
         \begin{tabular}{lcccccc}
            \hline
            \noalign{\smallskip}
              ${\rm M_{ini}}$ & ${\rm N_{TP}}$ & ${\rm M_{c}}$ & $<{\rm T_{HeBS}}>$ & $<{\rm T_{BCE}}>$ & ${\rm M^{tot}_{dup}}$ & $<{\rm \lambda}>$\\
               (\msun)  & & (\msun) & (MK) & (MK) & (\msun) &  \\
            \noalign{\smallskip}
            \hline
            \noalign{\smallskip}
3.0 & 18 & 0.83 & 306 & 38 & 0.05 & 0.82\\
3.2 (OV) & 33 & 0.84 & 308 & 44 & 0.09 & 0.78\\
3.5 & 23 & 0.88 &   299 &    43 &   0.08 & 0.69\\
            \noalign{\smallskip}
            \hline \noalign{\smallskip}
          \end{tabular}
          \caption{Summary of the main structure parameters of the models computed with \textsc{monstar}. ${\rm M_{ini}}$ is the initial mass, ${\rm N_{TP}}$ is the number of thermal pulses. ${\rm M_{c}}$ is the size of the H-exhausted core at the end of the evolution. $<{\rm T_{HeBS}}>$ and $<{\rm T_{BCE}}>$ are, respectively, the average temperatures at the bottom of the He-burning shell and the base of the convective envelope. ${\rm M^{tot}_{dup}}$ is the total mass dredged-up, and $<{\rm \lambda}>$ is the average of the dredge-up parameter.}
          \label{tab:evol1}
    \end{center}
\end{table}
\endgroup

\subsection{Nucleosynthesis code}
To calculate the detailed chemical yields of our models, the results from \textsc{monstar} were post-processed with the Monash nucleosynthesis code \textsc{monsoon} (e.g. \citealt{can93, Lattanzio1996,lugaro2012}), in particular the version described in \citealt{buntain2017}. The nuclear network includes 320 species and 2336 reactions \citep{lugaro2012}. Nuclear reaction rates are from the {\it Joint Institute for Nuclear Astrophysics} (JINA;  \citealt{cyb10}). The relevant reactions for the production of neutrons,  \iso{13}C($\alpha$,n)\iso{16}O and \iso{22}Ne($\alpha$,n)\iso{25}Mg are from \cite{heil2008} and \citealt{iliadis2010}, respectively. \iso{22}Ne($\alpha$,$\gamma$)\iso{26}Ne is also from the latter source. Neutron-capture reactions are from {\it KADoNiS} \citep{dillmann2006}.

\textsc{monsoon} allows for the introduction of a partially mixed zone (PMZ) through which envelope protons are forced to enter the intershell region at the time of the deepest advance of the convective envelope during the third dredge-up episode (see e.g. \citealt{karakas2010} for details). This approach \citep{straniero1995} leads to a `\iso{13}C-pocket' when the protons are captured by \iso{12}C, allowing the s-process to proceed through the \iso{13}C($\alpha$,n)\iso{16}O neutron source. In our calculations, we varied the size of the PMZ to test the effect on heavy element nucleosynthesis with values between $5\times 10^{-4}$ \msun{} and  $1\times 10^{-3}$ \msun{}, in agreement with \citealt{lugaro2012} for their low-metallicity ($\rm{Z}=0.0001$) models.

\begin{figure*}[h]
   \centering   \includegraphics[width=.78\linewidth]
      {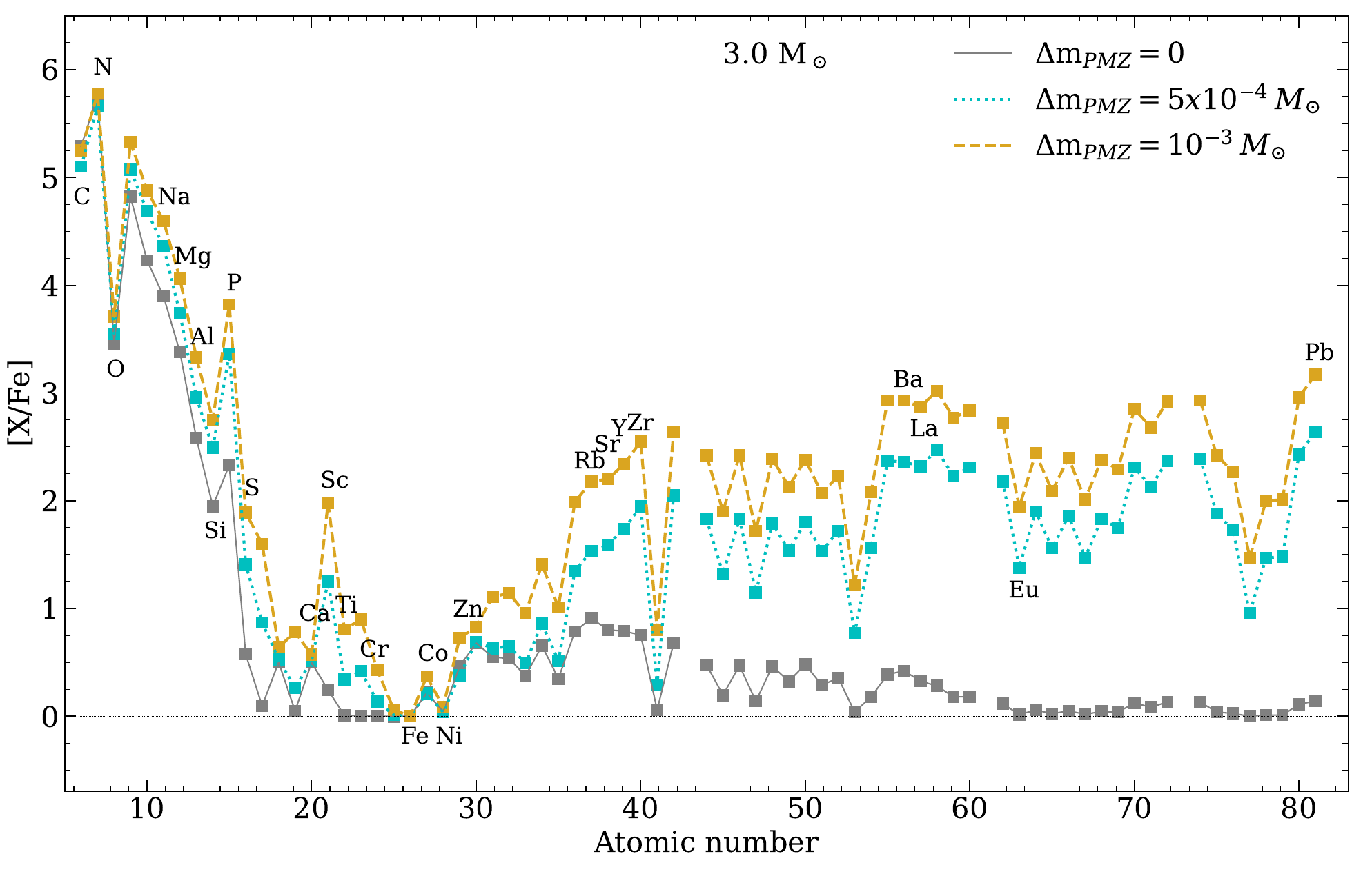}
         \caption{Yield abundances from the 3 \msun models computed without PMZ (grey), with $\Delta m_{PMZ}=5\times 10^{-4}$ \msun{} (cyan) and $\Delta m_{PMZ}=10^{-3}$ \msun{} (gold).}
         \label{fig:Comp_pockets}
   \end{figure*}
   
\subsection{Initial Composition}
\label{sec:initial_comp}
As we hypothesise that HE~1327 started with a typical chemical pattern of its epoch, we consider observations of the state of Galactic chemical evolution (GCE) at [Fe/H]~$=-5.2$. The study of \citealt{kobayashi2020} provides a very useful compilation of many elemental abundances down to very low [Fe/H]. Even though most data do not reach as low as [Fe/H]~$=-5$, the trends are usually clear. Primarily we see the well-known $\alpha$-enhancements $\simeq +0.5$~dex (e.g. \citealt{abia1989}). Of particular note for HE~1327 is Zn since it has been observed to be substantially super-solar \citep{ezzeddine2019}. It is well established that there is an increase of Zn towards lower metallicities, which has been explored in detail both observationally (\citealt{Saito2009}) and in models (\citealt{Umeda2002}). This trend strongly suggests that the Zn content of HE~1327 has not been enhanced above normal for its [Fe/H]. Quantitatively, the [Zn/Fe] vs [Fe/H] relation from \citealt{Saito2009} (their Eqn.~1) predicts $\rm{[Zn/Fe]} = +1.0$~dex at the metallicity of HE~1327. This is consistent with the observed [Zn/Fe]~$ = 0.80 \pm 0.25$~dex \citep{ezzeddine2019}. We adopt a conservative value of $+0.7$~dex for the initial value of our models. We note that Zn is hardly affected in low-mass stars.
Apart from the $\alpha$- and Zn-enhancements, our models start with scaled-solar composition \citep{grevesse1996} at a metallicity of [Fe/H]~$=-5.2$.

\subsection{Summary of evolution and nucleosynthetic yields}
   
\citealt[hereafter GP22]{gilpons2022} described the evolution and light-element nucleosynthesis in intermediate-mass primordial to extremely metal-poor stars. The yield of the 3.0 \msun\: solar-scaled composition $Z=10^{-7}$ model in GP22 showed a promising similarity to HE~1327 (their Fig.~11). This model experienced a very weak proton-ingestion episode (PIE) during the early thermally-pulsing asymptotic giant branch (TP-AGB) phase. However, $\alpha$-enhancement hampers the occurrence of these episodes in our current models, which are at the mass and metallicity boundary of PIE occurrence (see Fig.~4 of \citealt{campbell2008}). Otherwise, the evolution proceeds in a way very similar to that described in GP22.

The nucleosynthetic yields for our 3.0~\msun\ models are shown in Fig.~\ref{fig:Comp_pockets}. We include a model with no PMZ to isolate the elemental production from TDU+HBB (third dredge-up and hot-bottom-burning) from the effects of the \iso{13}C-pockets. The light elements, from C to P, show substantial enhancements ($\sim 2 - 7$~dex). The [N/Fe]>[C/Fe]>[O/Fe] pattern is very apparent, along with the light-element `slide' pattern, from F to Cl -- both patterns which are qualitatively similar to that of HE~1327. 

The light-element slide abundance pattern results from dredge-up each TP cycle. The hydrogen shell provides the elements produced/destroyed via proton capture (eg. Na, Mg, Al). Additionally, this material is subject to further nucleosynthesis through neutron captures. At such low metallicities, the $^{22}$Ne source is active at this mass, with temperatures in the He shell reaching 300~MK (Table~\ref{tab:evol1}). The $^{13}$C neutron source is also weakly active, becoming more dominant when a $^{13}$C pocket is added. Given these available neutrons, a chain of neutron captures starting from the abundant lighter elements ensues. For example, starting from \iso{27}Al (which is produced via \iso{26}Mg(p,$\gamma$)\iso{27}Al in the H-shell), the chain is essentially a light-element s-process, following closely the valley of beta stability: \iso{27}Al(n,$\gamma$)\iso{28}Al($\beta^{-}$)\iso{28}Si(n,$\gamma$)\iso{29}Si(n,$\gamma$)\iso{30}Si(n,$\gamma$)\iso{31}Si($\beta^{-}$)\iso{31}P, and so on. The result is the characteristic light-element abundance slide.

Hot bottom burning, at the base of the convective envelope, also occurs in our model (particularly during the later TPs). At around 40~MK, the HBB nucleosynthesis is not advanced, with its main effect being the conversion of dredged-up C to N via the CN cycle, giving rise to the characteristic CNO pattern dominated by N.

   \begin{figure*}[t]
   \centering \includegraphics[width=.8\linewidth]{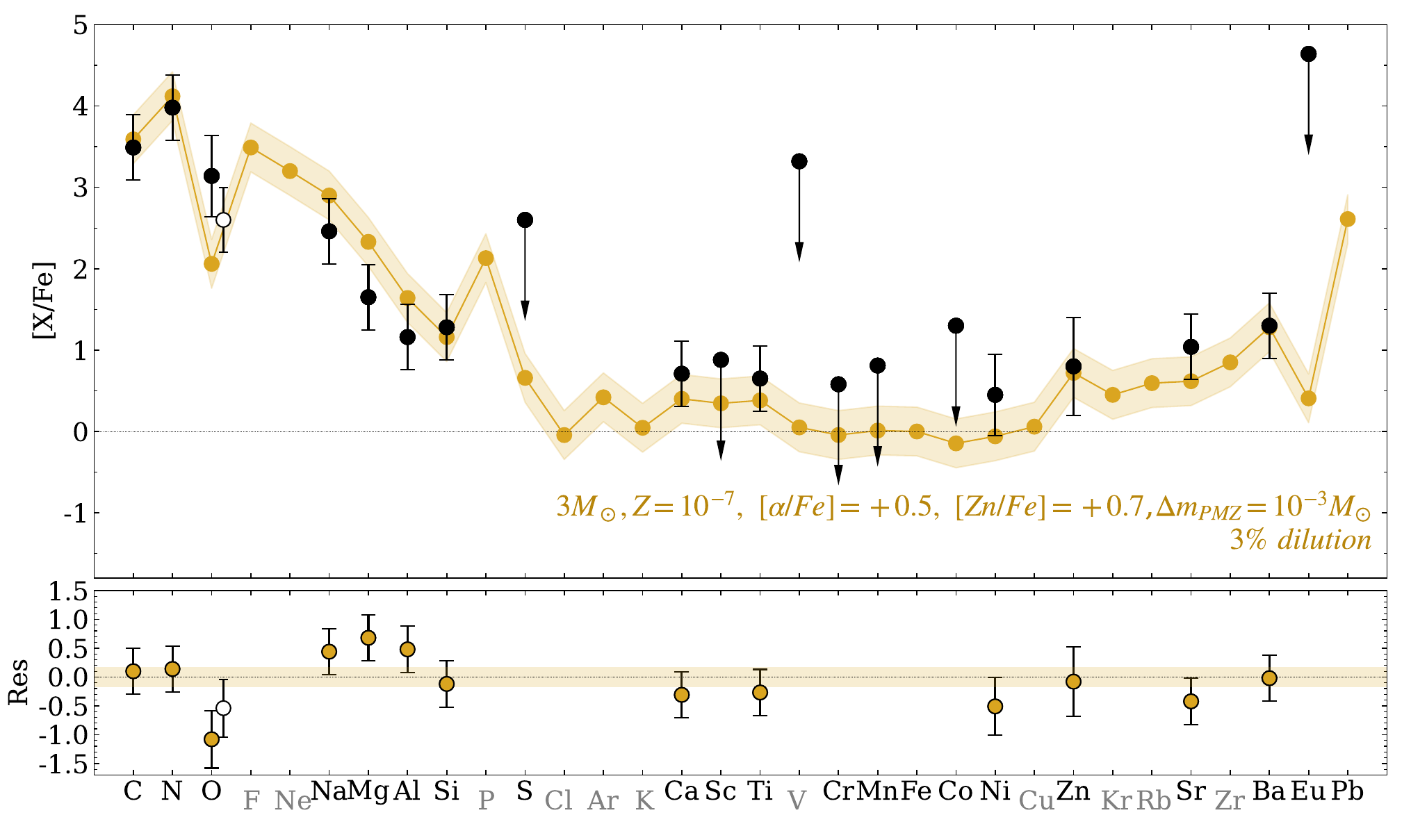}
      \caption{Upper panel: Diluted yield abundances from the 3 \msun{} model with $\Delta m_{PMZ}=10^{-3}$ (gold), observed abundances (black), and upper limits (grey). 
      Error bars are 2$\sigma$.   
      Colour shadings represent the approximate $\pm0.3$~dex uncertainty associated with abundances in stellar models (see main text and Appendix~\ref{appx:uncerts} for details).
      We show two determinations for O (solid circle from \citealt{ezzeddine2019}, and open circle from \citealt{frebel2006}), to highlight the uncertainty in measuring this element.
      Elements for which observations are unavailable are labelled in grey.
      Observed abundances are compiled from the literature; most are from Table~1 in \citealt{ezzeddine2019}, which incorporates the 3D values from \citealt{frebel2008}. Abundances for Si and Fe are from \citealt{ezzeddine2018}. The upper limit for S is from \citep{bonifacio2012}, and V and Cr from \citep{frebel2008}.
      Lower panel: Residuals from the comparison.}
         \label{fig:Comp_obs}
   \end{figure*}

It can be seen that introducing a PMZ has little effect on most of the light elemental yields. The exceptions are P, which has an additional +1.4~dex enhancement over the no-PMZ model, S (+1.4~dex), Cl (+1.4~dex), and Sc (+1.8~dex). 
Despite these significant enhancements, we note that due to the level of dilution required to match HE~1327 (a factor of $\sim 1.5$~dex; Sec.~\ref{sec:compare_ours}), enhancements over the initial composition of less than $\sim 1.5$~dex will be of no observational consequence. This is because the ISM content will dominate for those elements. Thus, for the light elements, we expect significant ($> 0.5$~dex) enhancements of only C to P in the final, diluted composition (as seen in Fig.~\ref{fig:Comp_obs}).

Phosphorus is particularly enhanced. As noted above, it is produced through a neutron capture chain starting with the lighter elements, with the final step to \iso{31}P being a $\beta^{-}$ decay after a neutron capture on \iso{30}Si (e.g. \citealt{karakas2010}). With the introduction of the PMZ and the associated increases in neutron density, this nucleosynthesis channel is enhanced. With a total enhancement of around $+4$~dex, this element may be observable and is thus a prediction of our model. We note however, that the exact production of \iso{31}P is dependent on the uncertain rate of \iso{30}Si(n,$\gamma$) (see e.g. \citealt{Fok2024}).

Moving to the heavy elements, our models without a PMZ do not develop an efficient s-process. As in higher metallicity cases at this mass (e.g. \citealt{fishlock2014}),  \iso{22}Ne($\alpha$,n)\iso{25}Mg is only marginally active as a neutron source during thermal pulses, resulting in moderate enhancements ($+0.7 \rightarrow +1.0$~dex) of [Kr/Fe], [Rb/Fe], [Sr/Fe], [Zr/Fe] and [Ba/Fe] in the yield. 
The introduction of a PMZ naturally leads to considerably higher neutron densities. Successive neutron captures on \iso{22}Ne (that is refilled in every TDU episode and only partially consumed during the NeNa-cycle) lead to the formation of \iso{56}Fe. This allows the s-process to continue, as described in \citealt{bisterzo2010},
producing enhancements of around two orders of magnitude between Kr and Ba. These abundances rise for increasingly heavier elements up to Pb (Fig.~\ref{fig:Comp_pockets}). Pb itself is particularly abundant, with [Pb/Fe] reaching around $+4$~dex ($\sim +2$~dex on dilution, see next section and Fig.~\ref{fig:Comp_obs}). Pb has not yet been observed in HE~1327 and is another key prediction of our model.

Finally, our model also predicts that many elements will \textit{not} be significantly enhanced: S, Cl, K, V, Cu, and Eu (see Fig~\ref{fig:Comp_obs}). This is another useful observational check for our scenario.

\section{Comparison to observations} \label{sec:compareobs}

\subsection{Intermediate-mass star Scenario: Our models}
\label{sec:compare_ours}

As described in Section~\ref{sec:initial_comp}, our scenario assumes that some chemical evolution had already occurred by $\rm{[Fe/H]} = -5.2$, giving rise to star-forming clouds with typical enhancements of [$\alpha$/Fe]~$=+0.5$ and [Zn/Fe]~$=+0.7$~dex (Sec.~\ref{sec:models}). From here, one of two (similar) scenarios played out: (i) an intermediate-mass star formed and polluted the interstellar medium, and HE~1327 formed from that gas, or (ii) both HE~1327 and the intermediate-mass star formed at the same time in a binary system, and the intermediate-mass star polluted HE~1327 once it became an AGB star. 
In either scenario, the chemical pattern of HE~1327 is explained by a superposition of intermediate-mass star composition onto the natal composition. As the lifetime of a 3.0~\msun~star is 230 Myr, this sets the timescale of the scenarios. We note that both scenarios have roughly the same timescale because it is mainly dependent on the main sequence lifetime of the star, before it reaches the AGB phase (the AGB phase is relatively short-lived). Interestingly, this is a shorter timescale than that reported from recent JWST observations, which show metals and unusually high N/O ratios very early in the Universe (440~Myr post-Big Bang; \citealt{Bunker2023,Cameron2023}). This lends weight to the suggestion that intermediate-mass stars may have significantly contributed to the ISM at that time. We note that rapidly-rotating massive and very massive models could also give rise to the high N/O ratios (e.g. \citealt{Nandal2024,Tsiatsiou2024}).

We diluted the intermediate-mass star yields into the natal gas to mimic the ISM pollution scenario. We find that the ejecta from our intermediate-mass star must be mixed in $\sim$100~\msun{} of interstellar medium matter. A dilution factor f$_{dil}=3\%$ of intermediate-mass star ejecta within the natal cloud provides the best overall fit to observations.
Figure~\ref{fig:Comp_ours_obs} shows observed and theoretical abundances from some selected models. In this figure, and Figure~\ref{fig:Comp_obs}, the shaded regions around the model yield data represent an attempt to highlight their associated uncertainties. In Appendix~\ref{appx:uncerts} we detail why we use 0.3~dex as an approximate abundance error. 
The models in Figure~\ref{fig:Comp_ours_obs} generally fit within uncertainties, except for the underproduction of O. 
Some models overproduce Mg. Sr and Ba form with relative abundances that agree with observations ([Ba/Fe] > [Sr/Fe]), and their fit is better for increasingly larger PMZ sizes.   
The results from the 3.5~\msun{} model (with more efficient HBB than the lower mass cases) with $\Delta m_{PMZ}=5\times 10^{-4}$ \msun{}\: are characterised by final [C/N] values lower than those from the 3.0~\msun{} models, and Na, Mg and Al are overproduced.
The introduction of overshooting in the 3.2 \msun{} model yields results similar to those of the 3.0~\msun{} models without overshooting.

Figure \ref{fig:Comp_obs} shows our best fit to the  observations of HE~1327: the 3.0~\msun{} model with no overshooting and $\Delta m_{PMZ}=10^{-3}$~\msun{}. All abundances match within the uncertainties except for oxygen, which is $0.5 - 1.0$~dex underproduced. We stress that we have not excluded any (measured) elements in our fits. The model is also consistent with all seven upper limits.

\begin{figure*}[h]
    \centering
\includegraphics[width=\linewidth]{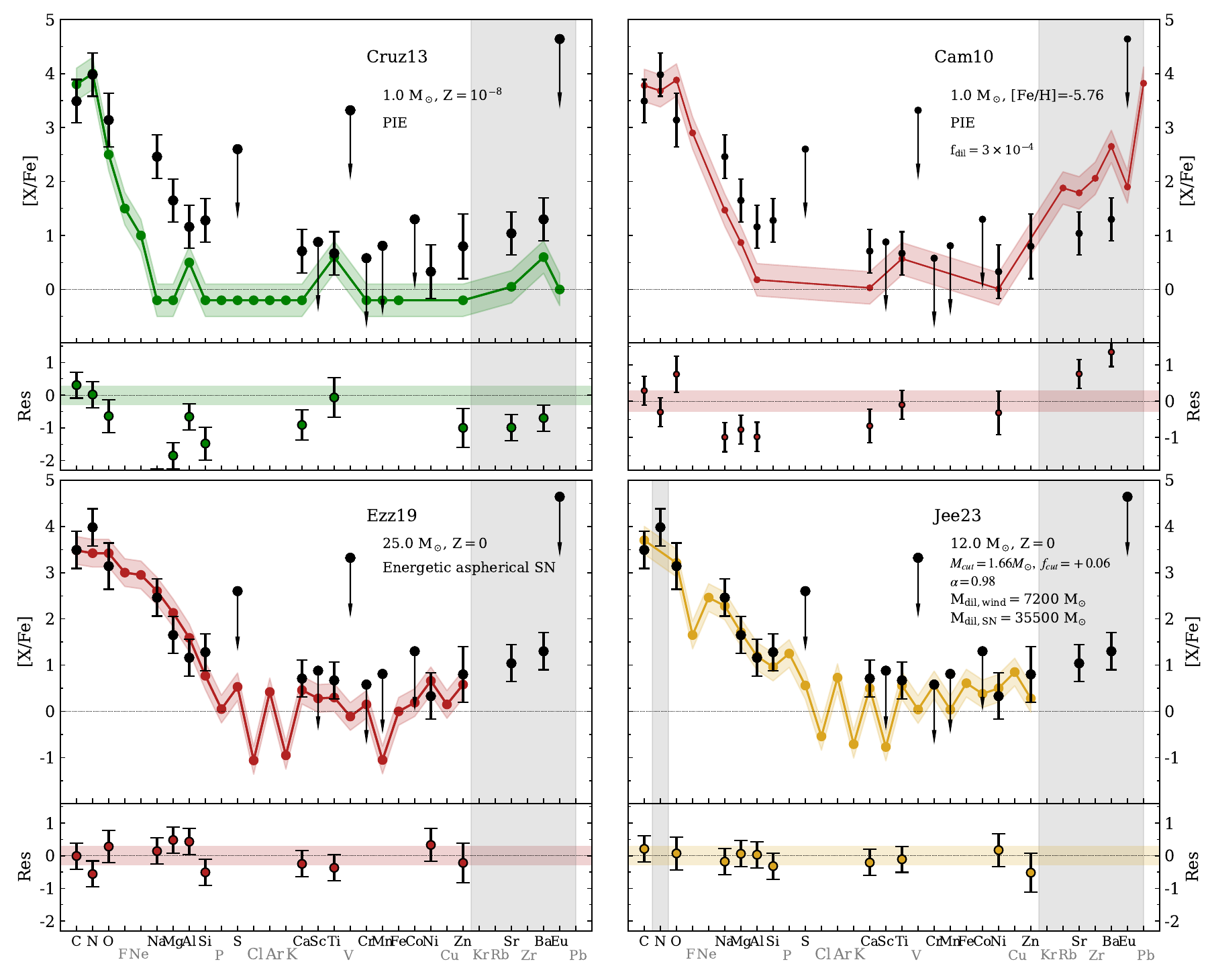}
      \caption{Observed abundances of HE~1327 and observational upper limits (black) together with the yield abundances from different stellar models.
      Upper panels show results from 1\msun{} models undergoing PIEs, and lower panels show results from massive star models undergoing supernova explosions. The $Z=10^{-8}$ model (upper left) is from \cite{cruz13}. The [Fe/H]~$=-5.76$ (upper right) is from \cite{campbell2010}.  
      The yields from \cite{ezzeddine2019} (lower left panel) are from a primordial 25 \msun star undergoing an asymmetric mixing and fallback supernova explosion. 
      The yields from \cite{jeena2023} (lower right panel) correspond to a rapidly rotating 12 \msun star of Z=0, undergoing very efficient mixing and quasi-chemically homogeneous evolution prior to a supernova explosion. 
      The grey-shaded areas highlight the elements for which no theoretical abundances were reported. Note that in the model by \citealt{jeena2023}, the authors did not consider separately the abundances of C and N, but added the abundances of the two elements, and plotted the corresponding value at the location of C. We added the grey bar at the location of N to note this.
      As in Figure \ref{fig:Comp_obs}, error bars are 2$\sigma$, and shadings have the same meaning. 
      We present two determinations for O (solid circle from \citealt{ezzeddine2019} and open circle from \citealt{frebel2006}) to highlight the uncertainty in measuring this element.
      Elements for which observations are unavailable are labelled in grey.
      Observed abundances are compiled from the literature; most are from Table~1 in \citealt{ezzeddine2019}, which incorporates the 3D values from \citealt{frebel2008}. Abundances for Si and Fe are from \citealt{ezzeddine2018}. The upper limit for S is from \citep{bonifacio2012}, and V and Cr from \citep{frebel2008}.
      Lower panel: Residuals from the comparison. The $\chi^2$ values are given in Table~\ref{tab:chi2}.}
         \label{fig:Comp_other}
\end{figure*}

Oxygen, the one element in our model that does not fit, is challenging to measure since it suffers from substantial NLTE and 3D effects. In Figure \ref{fig:Comp_obs}, we show two O abundance measurements to highlight this, from \citealt{ezzeddine2019} and \citealt{frebel2006}\footnote{The first observations of HE~1327 \citep{fre05,aoki2006} only provided upper thresholds for [O/Fe] ($<4.0$~dex). \cite{frebel2006} revisited 1D-LTE abundances with VLT/UVES including 3D-1D corrections \citep{asplund2005}, and estimated a correction for 3D effects of $-0.9$~dex, reducing [O/Fe] from +3.7 to +2.8. \cite{frebel2008} confirmed the 1D abundances at higher S/N and proposed a more physically sound 3D-correction technique. This resulted in a smaller 3D correction leading to $\rm{[O/Fe]}=+3.4$~dex. The later [Fe/H] correction in \cite{ezzeddine2018} led to the currently recommended $\rm{[O/Fe]}=+3.14$~dex.}.
We note that model uncertainties could also cause this level of disagreement ($0.5-1.0$~dex in [O/Fe]). For instance, increased oxygen could come from deeper (or more numerous) TDU episodes, which depend on the poorly known convective boundaries and wind efficiency. Mixing at the bottom of the convective He flash zone may also increase surface oxygen in AGB models \citep{herwig2004}  -- without significantly altering the C abundance, given the efficiency of hot-bottom burning in transforming this element into N.

\subsection{Previous scenarios: Low-mass PIEs and Supernovae}
\label{sec:compare_pie}

As detailed in Section~\ref{sec:intro}, two main scenarios have so far been considered in the literature -- SNe and PIEs. Here we provide quantitative comparisons between observations and models for these scenarios, 
using the $\chi^2$-test, as in \citealt{bisterzo2011}. In this way, we aim to see which model(s) fit best. We provide further details of these tests in Appendix~\ref{app:details}, but report the main findings here.


As discussed in the Introduction, the low-mass star PIE scenario has not met with much success in explaining HE~1327. This can be seen in panels 1 and 2 of Figure~\ref{fig:Comp_other}, where we compare the PIE models of \cite{campbell2010} and \cite{cruz13} with the observed composition of HE~1327. The 1~\msun{}, [Fe/H]~$=-6.5$ model in \cite{campbell2010}  does not reproduce the [N/Fe]>[C/Fe]>[O/Fe] pattern, overproduces the light-element slide, and severely overproduces Ba.
The 1~\msun{}, $Z=10^{-8}$ model of \cite{cruz13} could reproduce [Sr/Fe] in HE~1327, but only at the expense of overproducing C, N and O by more than two orders of magnitude. As can be seen in Figure~\ref{fig:Comp_other}, when CNO is fit, the model fails to match Na, Mg and Si, as well as Zn and Sr.

Comparisons of some representative SN model chemical patterns against observed abundances are shown in the lower panels of Figure \ref{fig:Comp_other}. As mentioned in the Introduction, SN models have had more success in matching the abundance pattern of HE~1327. It should be noted, however, that usually not all observed elements were part of the matches (particularly Sr) -- we highlight this with grey shading in the figures.

The best fit from the aspherical bipolar jet SN model of \citealt{ezzeddine2019} (panel 3 of Fig.~\ref{fig:Comp_other}) matches all observed elements except Sr (which was not reported; Ba had not been observed yet). As mentioned in Section~\ref{sec:intro}, \citealt{ezzeddine2019}  suggested that Sr could come from fast-rotating Pop III supernova progenitors.
The CNO pattern, where N is highest in HE 1327, was not matched by this model  -- although we suggest N is within the model and observational uncertainties. Similarly to the \citealt{ezzeddine2019} model, the rapidly-rotating model of \citealt{jeena2023} (Section~\ref{sec:intro} and panel 4 of Fig.~\ref{fig:Comp_other})  matches all of the elements except Sr (not reported). It has the same issue as the other massive star/SN models in that the N>C>O pattern is not reproduced. In addition, Zn is only marginally reproduced.

For completeness, we briefly describe some earlier SN models for HE~1327. The model of \cite{iwamoto2005}, a 25~\msun{} faint ($E<10^{51}\:erg$) core-collapse SN with mixing and fallback, reproduced the Na, Mg, Al slide and the Ca abundance, but failed to explain the [N/Fe]>[C/Fe]>[O/Fe] pattern and the formation of Ni, Zn, Sr and Ba.
\citealt{takahashi2014} fitted HE~1327 with the yields from a low-energy rotating SN of 20 \msun{}. This model reproduced C, O and the light-element slide but significantly underproduced N and Si. No elements heavier than Si were reported.

Finally, we considered the recent hyper–metal-poor and primordial models of \citet{roberti2024} (also see \citealt{frischknecht2012,frischknecht2016}). These are fast-rotating models that include heavy elements. As noted earlier, fast-rotating models may produce the Sr and Ba seen in HE~1327. However, the yields of these models differ strongly from the observed abundance pattern of HE~1327. Neither of the three characteristic patterns of the star is reproduced: neither the CNO pattern, nor the light-element slide, nor the heavy-elements. The main reason for this discrepancy is probably that these models do not implement the mixing–fallback mechanism to compute the explosive yields \citep{maeda2003,umeda2003}, an explosion model originally proposed to reproduce the chemical patterns of extremely metal-poor stars, and subsequently applied in various forms in the literature (for instance, in the SN models that we present in Figure \ref{fig:Comp_other}, and analyse in Appendix \ref{app:details}). While primordial and hyper–metal-poor fast-rotating stars may still play a role in Galactic chemical evolution or in shaping the abundance patterns of many CEMP stars, the current fast-rotating models considered in the context of a single-polluter scenario fail to account for the observed composition of HE~1327.

In summary, for the Pop~III SN scenario, some of the fits have been quite successful, although they have systematically disregarded the detection of [Sr/Fe$] \approx +1$~dex, and have not been able to match the CNO pattern reliably.
We note in the context of these massive star/SN models that the observed [Ba/Sr$]\simeq +0.2$~dex suggests the operation of the main s-process (or possibly r-process) and likely rules out the weak s-process expected from massive stars. 

Overall, we find that intermediate-mass models perform comparably to or even better than the best-fitting supernova models, especially when all the observed species are considered (see $\chi^2$ Case A in Table~\ref{tab:chi2}). 

\section{Summary and Conclusion}\label{sec:discuss}

Although often categorised as a CEMP-no star, we argued that, given its enhanced [N/Fe], [Sr/Fe] and (recently reported) high [Ba/Fe], HE~1327 should be categorised as a N-CEMP-s star -- and that the question of its origin should be revisited.

We proposed that the composition pattern of HE~1327 is a result of a superposition of a chemical yield from an intermediate-mass AGB star ($\sim 3$~\msun{}) onto a primitive cloud, which had a typical composition pattern of the time. This superposition could have occurred either through binary mass-transfer or pre-pollution of HE~1327's formation cloud by the AGB star. Interestingly, we found that the high Zn content (and $\alpha$-element content) of HE~1327 can be explained by Galactic chemical evolution -- Zn is observed to increase at the lowest metallicties.
We tested our scenario by comparing the observed abundances of HE~1327 to our theoretical yields of intermediate-mass stars diluted with primitive gas cloud material.

Our best-fitting model reproduces simultaneously the full range of \emph{light and heavy} elements observed in HE~1327. The exception is O, which is underproduced, but we noted the difficulties in determining its observed abundance and the uncertainty in stellar models. We emphasise that, unlike SN models in the literature aimed to explain the composition pattern of HE~1327, our model also matches Sr and Ba, and does not require exotic stellar evolution scenarios, just standard intermediate-mass star evolution.   

We note, however, that HE~1327 is an atypical star at this metallicity. It stands out due to its s-process enrichment, which is not seen in other stars of comparable metallicity. Thus, our AGB plus early Galactic chemical evolution scenario may not apply more generally at such early times.

Our models also point to a possible way to differentiate between massive star pollution and intermediate-mass star pollution -- by observing P and Pb. Intermediate-mass star models with a PMZ give [P/Fe] and [Pb/Fe] significantly higher than those from standard SN models. That said, alternative models of massive stars involving rotation and nucleosynthesis in convective-reactive regions might also explain the existence of P \citep{masseron2020}.

In conclusion, our results imply that, at metallicity $\rm{[Fe/H]}\sim -5.2$, the interstellar medium may have already shown some degree of homogeneity, that intermediate-mass stars formed and polluted this environment, and that we have evidence from these pioneering stars in HE~1327. We note that models of Pop III star formation suggest that intermediate-mass stars could have formed very early (e.g. \citealt{Sharda2021}) and that recent observations from JWST show metals and an unusually high N/O ratio just 440 Myr after the Big Bang \citep{Bunker2023,Cameron2023}. Although this composition has so far been attributed to rotating massive stars, this timescale is also compatible with the $\sim 230$~Myr lifetimes of our ancient stars of around  3~\msun, which produce high concentrations of N as well.

\begin{acknowledgements}
     This work was supported by the Spanish project PID2019-109363GB-100 and  by the German \emph{Deut\-sche For\-schungs\-ge\-mein\-schaft, DFG\/} project number Ts~17/2--1. SWC acknowledges federal funding from the Australian Research Council through a Future Fellowship (FT160100046) and Discovery Projects (DP190102431 \& DP210101299). Parts of this research was supported by the Australian Research Council Centre of Excellence for All Sky Astrophysics in 3 Dimensions (ASTRO 3D), through project number CE170100013. This research was supported by use of the Nectar Research Cloud, a collaborative Australian research platform supported by the National Collaborative Research Infrastructure Strategy (NCRIS).
\end{acknowledgements}

%
   \bibliographystyle{aa} 
   \bibliography{AA_GP2025_final.bib} 

\begin{thebibliography}{86}
\expandafter\ifx\csname natexlab\endcsname\relax\def\natexlab#1{#1}\fi

\bibitem[{{Abia} \& {Rebolo}(1989)}]{abia1989}
{Abia}, C. \& {Rebolo}, R. 1989, \apj, 347, 186

\bibitem[{{Aoki} {et~al.}(2006){Aoki}, {Frebel}, {Christlieb}, {Norris},
  {Beers}, {Minezaki}, {Barklem}, {Honda}, {Takada-Hidai}, {Asplund}, {Ryan},
  {Tsangarides}, {Eriksson}, {Steinhauer}, {Deliyannis}, {Nomoto}, {Fujimoto},
  {Ando}, {Yoshii}, \& {Kajino}}]{aoki2006}
{Aoki}, W., {Frebel}, A., {Christlieb}, N., {et~al.} 2006, \apj, 639, 897

\bibitem[{{Asplund}(2005)}]{asplund2005}
{Asplund}, M. 2005, \araa, 43, 481

\bibitem[{{Beers} \& {Christlieb}(2005)}]{beers2005}
{Beers}, T.~C. \& {Christlieb}, N. 2005, \araa, 43, 531

\bibitem[{{Beyer}(1991)}]{standardmat1991}
{Beyer}, W.~H. 1991, {CRC standard mathematical tables and formulae} (CRC
  Press)

\bibitem[{{Bisterzo} {et~al.}(2010){Bisterzo}, {Gallino}, {Straniero},
  {Cristallo}, \& {K{\"a}ppeler}}]{bisterzo2010}
{Bisterzo}, S., {Gallino}, R., {Straniero}, O., {Cristallo}, S., \&
  {K{\"a}ppeler}, F. 2010, \mnras, 404, 1529

\bibitem[{{Bisterzo} {et~al.}(2011){Bisterzo}, {Gallino}, {Straniero},
  {Cristallo}, \& {K{\"a}ppeler}}]{bisterzo2011}
{Bisterzo}, S., {Gallino}, R., {Straniero}, O., {Cristallo}, S., \&
  {K{\"a}ppeler}, F. 2011, \mnras, 418, 284

\bibitem[{{Bloecker}(1995)}]{bloecker1995}
{Bloecker}, T. 1995, \aap, 297, 727

\bibitem[{{B{\"o}hm-Vitense}(1958)}]{bom58}
{B{\"o}hm-Vitense}, E. 1958, \zap, 46, 108

\bibitem[{{Bonifacio} {et~al.}(2012){Bonifacio}, {Caffau}, {Venn}, \&
  {Lambert}}]{bonifacio2012}
{Bonifacio}, P., {Caffau}, E., {Venn}, K.~A., \& {Lambert}, D.~L. 2012, \aap,
  544, A102

\bibitem[{{Bunker} {et~al.}(2023){Bunker}, {Saxena}, {Cameron}, {Willott},
  {Curtis-Lake}, {Jakobsen}, {Carniani}, {Smit}, {Maiolino}, {Witstok},
  {Curti}, {D'Eugenio}, {Jones}, {Ferruit}, {Arribas}, {Charlot}, {Chevallard},
  {Giardino}, {de Graaff}, {Looser}, {L{\"u}tzgendorf}, {Maseda}, {Rawle},
  {Rix}, {Del Pino}, {Alberts}, {Egami}, {Eisenstein}, {Endsley}, {Hainline},
  {Hausen}, {Johnson}, {Rieke}, {Rieke}, {Robertson}, {Shivaei}, {Stark},
  {Sun}, {Tacchella}, {Tang}, {Williams}, {Willmer}, {Baker}, {Baum},
  {Bhatawdekar}, {Bowler}, {Boyett}, {Chen}, {Circosta}, {Helton}, {Ji},
  {Kumari}, {Lyu}, {Nelson}, {Parlanti}, {Perna}, {Sandles}, {Scholtz},
  {Suess}, {Topping}, {{\"U}bler}, {Wallace}, \& {Whitler}}]{Bunker2023}
{Bunker}, A.~J., {Saxena}, A., {Cameron}, A.~J., {et~al.} 2023, \aap, 677, A88

\bibitem[{{Buntain} {et~al.}(2017){Buntain}, {Doherty}, {Lugaro}, {Lattanzio},
  {Stancliffe}, \& {Karakas}}]{buntain2017}
{Buntain}, J.~F., {Doherty}, C.~L., {Lugaro}, M., {et~al.} 2017, \mnras, 471,
  824

\bibitem[{{Cameron} {et~al.}(2023){Cameron}, {Katz}, {Rey}, \&
  {Saxena}}]{Cameron2023}
{Cameron}, A.~J., {Katz}, H., {Rey}, M.~P., \& {Saxena}, A. 2023, \mnras, 523,
  3516

\bibitem[{{Campbell}(2008)}]{campbell_phd}
{Campbell}, S.~W. 2008, PhD Thesis, Monash University, Australia,
  arXiv:2410.21972

\bibitem[{{Campbell} \& {Lattanzio}(2008)}]{campbell2008}
{Campbell}, S.~W. \& {Lattanzio}, J.~C. 2008, \aap, 490, 769

\bibitem[{{Campbell} {et~al.}(2010){Campbell}, {Lugaro}, \&
  {Karakas}}]{campbell2010}
{Campbell}, S.~W., {Lugaro}, M., \& {Karakas}, A.~I. 2010, \aap, 522, L6

\bibitem[{{Cannon}(1993)}]{can93}
{Cannon}, R.~C. 1993, \mnras, 263, 817

\bibitem[{{Choplin} {et~al.}(2017){Choplin}, {Hirschi}, {Meynet}, \&
  {Ekstr{\"o}m}}]{choplin2017}
{Choplin}, A., {Hirschi}, R., {Meynet}, G., \& {Ekstr{\"o}m}, S. 2017, \aap,
  607, L3

\bibitem[{{Collet} {et~al.}(2006){Collet}, {Asplund}, \&
  {Trampedach}}]{collet2006}
{Collet}, R., {Asplund}, M., \& {Trampedach}, R. 2006, \apjl, 644, L121

\bibitem[{{Constantino} {et~al.}(2014){Constantino}, {Campbell}, {Gil-Pons}, \&
  {Lattanzio}}]{constantino2014}
{Constantino}, T., {Campbell}, S., {Gil-Pons}, P., \& {Lattanzio}, J. 2014,
  \apj, 784, 56

\bibitem[{{Constantino} {et~al.}(2015){Constantino}, {Campbell},
  {Christensen-Dalsgaard}, {Lattanzio}, \& {Stello}}]{Constantino2015}
{Constantino}, T., {Campbell}, S.~W., {Christensen-Dalsgaard}, J., {Lattanzio},
  J.~C., \& {Stello}, D. 2015, \mnras, 452, 123

\bibitem[{{Cruz} {et~al.}(2013){Cruz}, {Serenelli}, \& {Weiss}}]{cruz13}
{Cruz}, M.~A., {Serenelli}, A., \& {Weiss}, A. 2013, \aap, 559, A4

\bibitem[{{Cyburt} {et~al.}(2010){Cyburt}, {Amthor}, {Ferguson}, {Meisel},
  {Smith}, {Warren}, {Heger}, {Hoffman}, {Rauscher}, {Sakharuk}, {Schatz},
  {Thielemann}, \& {Wiescher}}]{cyb10}
{Cyburt}, R.~H., {Amthor}, A.~M., {Ferguson}, R., {et~al.} 2010, \apjs, 189,
  240

\bibitem[{{Dillmann} {et~al.}(2006){Dillmann}, {Heil}, {K{\"a}ppeler}, {Plag},
  {Rauscher}, \& {Thielemann}}]{dillmann2006}
{Dillmann}, I., {Heil}, M., {K{\"a}ppeler}, F., {et~al.} 2006, in American
  Institute of Physics Conference Series, Vol. 819, Capture Gamma-Ray
  Spectroscopy and Related Topics, ed. A.~{Woehr} \& A.~{Aprahamian}, 123--127

\bibitem[{{Doherty} {et~al.}(2010){Doherty}, {Siess}, {Lattanzio}, \&
  {Gil-Pons}}]{doherty2010}
{Doherty}, C.~L., {Siess}, L., {Lattanzio}, J.~C., \& {Gil-Pons}, P. 2010,
  \mnras, 401, 1453

\bibitem[{{Ezzeddine} \& {Frebel}(2018)}]{ezzeddine2018}
{Ezzeddine}, R. \& {Frebel}, A. 2018, \apj, 863, 168

\bibitem[{{Ezzeddine} {et~al.}(2019){Ezzeddine}, {Frebel}, {Roederer},
  {Tominaga}, {Tumlinson}, {Ishigaki}, {Nomoto}, {Placco}, \&
  {Aoki}}]{ezzeddine2019}
{Ezzeddine}, R., {Frebel}, A., {Roederer}, I.~U., {et~al.} 2019, \apj, 876, 97

\bibitem[{{Fishlock} {et~al.}(2014){Fishlock}, {Karakas}, {Lugaro}, \&
  {Yong}}]{fishlock2014}
{Fishlock}, C.~K., {Karakas}, A.~I., {Lugaro}, M., \& {Yong}, D. 2014, \apj,
  797, 44

\bibitem[{{Fok} {et~al.}(2024){Fok}, {Pignatari}, {C{\^o}t{\'e}}, \&
  {Trappitsch}}]{Fok2024}
{Fok}, H.~K., {Pignatari}, M., {C{\^o}t{\'e}}, B., \& {Trappitsch}, R. 2024,
  \apjl, 977, L24

\bibitem[{{Frebel} {et~al.}(2005{\natexlab{a}}){Frebel}, {Aoki}, {Christlieb},
  {Ando}, {Asplund}, {Barklem}, {Beers}, {Eriksson}, {Fechner}, {Fujimoto},
  {Honda}, {Kajino}, {Minezaki}, {Nomoto}, {Norris}, {Ryan}, {Takada-Hidai},
  {Tsangarides}, \& {Yoshii}}]{frebel2005}
{Frebel}, A., {Aoki}, W., {Christlieb}, N., {et~al.} 2005{\natexlab{a}}, \nat,
  434, 871

\bibitem[{{Frebel} {et~al.}(2005{\natexlab{b}}){Frebel}, {Aoki}, {Christlieb},
  {Ando}, {Asplund}, {Barklem}, {Beers}, {Eriksson}, {Fechner}, {Fujimoto},
  {Honda}, {Kajino}, {Minezaki}, {Nomoto}, {Norris}, {Ryan}, {Takada-Hidai},
  {Tsangarides}, \& {Yoshii}}]{fre05}
{Frebel}, A., {Aoki}, W., {Christlieb}, N., {et~al.} 2005{\natexlab{b}}, \nat,
  434, 871

\bibitem[{{Frebel} {et~al.}(2006){Frebel}, {Christlieb}, {Norris}, {Aoki}, \&
  {Asplund}}]{frebel2006}
{Frebel}, A., {Christlieb}, N., {Norris}, J.~E., {Aoki}, W., \& {Asplund}, M.
  2006, \apjl, 638, L17

\bibitem[{{Frebel} {et~al.}(2008){Frebel}, {Collet}, {Eriksson}, {Christlieb},
  \& {Aoki}}]{frebel2008}
{Frebel}, A., {Collet}, R., {Eriksson}, K., {Christlieb}, N., \& {Aoki}, W.
  2008, \apj, 684, 588

\bibitem[{{Frischknecht} {et~al.}(2016){Frischknecht}, {Hirschi}, {Pignatari},
  {Maeder}, {Meynet}, {Chiappini}, {Thielemann}, {Rauscher}, {Georgy}, \&
  {Ekstr{\"o}m}}]{frischknecht2016}
{Frischknecht}, U., {Hirschi}, R., {Pignatari}, M., {et~al.} 2016, \mnras, 456,
  1803

\bibitem[{{Frischknecht} {et~al.}(2012){Frischknecht}, {Hirschi}, \&
  {Thielemann}}]{frischknecht2012}
{Frischknecht}, U., {Hirschi}, R., \& {Thielemann}, F.-K. 2012, \aap, 538, L2

\bibitem[{{Frost} \& {Lattanzio}(1996)}]{frost1996}
{Frost}, C.~A. \& {Lattanzio}, J.~C. 1996, \apj, 473, 383

\bibitem[{{Gaia Collaboration} {et~al.}(2018){Gaia Collaboration}, {Brown},
  {Vallenari}, {Prusti}, {de Bruijne}, {Babusiaux}, {Bailer-Jones}, {Biermann},
  {Evans}, {Eyer}, {Jansen}, {Jordi}, {Klioner}, {Lammers}, {Lindegren},
  {Luri}, {Mignard}, {Panem}, {Pourbaix}, {Randich}, {Sartoretti}, {Siddiqui},
  {Soubiran}, {van Leeuwen}, {Walton}, {Arenou}, {Bastian}, {Cropper},
  {Drimmel}, {Katz}, {Lattanzi}, {Bakker}, {Cacciari}, {Casta{\~n}eda},
  {Chaoul}, {Cheek}, {De Angeli}, {Fabricius}, {Guerra}, {Holl}, {Masana},
  {Messineo}, {Mowlavi}, {Nienartowicz}, {Panuzzo}, {Portell}, {Riello},
  {Seabroke}, {Tanga}, {Th{\'e}venin}, {Gracia-Abril}, {Comoretto},
  {Garcia-Reinaldos}, {Teyssier}, {Altmann}, {Andrae}, {Audard},
  {Bellas-Velidis}, {Benson}, {Berthier}, {Blomme}, {Burgess}, {Busso},
  {Carry}, {Cellino}, {Clementini}, {Clotet}, {Creevey}, {Davidson}, {De
  Ridder}, {Delchambre}, {Dell'Oro}, {Ducourant},
  {Fern{\'a}ndez-Hern{\'a}ndez}, {Fouesneau}, {Fr{\'e}mat}, {Galluccio},
  {Garc{\'\i}a-Torres}, {Gonz{\'a}lez-N{\'u}{\~n}ez}, {Gonz{\'a}lez-Vidal},
  {Gosset}, {Guy}, {Halbwachs}, {Hambly}, {Harrison}, {Hern{\'a}ndez},
  {Hestroffer}, {Hodgkin}, {Hutton}, {Jasniewicz}, {Jean-Antoine-Piccolo},
  {Jordan}, {Korn}, {Krone-Martins}, {Lanzafame}, {Lebzelter}, {L{\"o}ffler},
  {Manteiga}, {Marrese}, {Mart{\'\i}n-Fleitas}, {Moitinho}, {Mora}, {Muinonen},
  {Osinde}, {Pancino}, {Pauwels}, {Petit}, {Recio-Blanco}, {Richards},
  {Rimoldini}, {Robin}, {Sarro}, {Siopis}, {Smith}, {Sozzetti}, {S{\"u}veges},
  {Torra}, {van Reeven}, {Abbas}, {Abreu Aramburu}, {Accart}, {Aerts},
  {Altavilla}, {{\'A}lvarez}, {Alvarez}, {Alves}, {Anderson}, {Andrei},
  {Anglada Varela}, {Antiche}, {Antoja}, {Arcay}, {Astraatmadja}, {Bach},
  {Baker}, {Balaguer-N{\'u}{\~n}ez}, {Balm}, {Barache}, {Barata}, {Barbato},
  {Barblan}, {Barklem}, {Barrado}, {Barros}, {Barstow}, {Bartholom{\'e}
  Mu{\~n}oz}, {Bassilana}, {Becciani}, {Bellazzini}, {Berihuete}, {Bertone},
  {Bianchi}, {Bienaym{\'e}}, {Blanco-Cuaresma}, {Boch}, {Boeche}, {Bombrun},
  {Borrachero}, {Bossini}, {Bouquillon}, {Bourda}, {Bragaglia}, {Bramante},
  {Breddels}, {Bressan}, {Brouillet}, {Br{\"u}semeister}, {Brugaletta},
  {Bucciarelli}, {Burlacu}, {Busonero}, {Butkevich}, {Buzzi}, {Caffau},
  {Cancelliere}, {Cannizzaro}, {Cantat-Gaudin}, {Carballo}, {Carlucci},
  {Carrasco}, {Casamiquela}, {Castellani}, {Castro-Ginard}, {Charlot},
  {Chemin}, {Chiavassa}, {Cocozza}, {Costigan}, {Cowell}, {Crifo}, {Crosta},
  {Crowley}, {Cuypers}, {Dafonte}, {Damerdji}, {Dapergolas}, {David}, {David},
  {de Laverny}, {De Luise}, {De March}, {de Martino}, {de Souza}, {de Torres},
  {Debosscher}, {del Pozo}, {Delbo}, {Delgado}, {Delgado}, {Di Matteo},
  {Diakite}, {Diener}, {Distefano}, {Dolding}, {Drazinos}, {Dur{\'a}n},
  {Edvardsson}, {Enke}, {Eriksson}, {Esquej}, {Eynard Bontemps}, {Fabre},
  {Fabrizio}, {Faigler}, {Falc{\~a}o}, {Farr{\`a}s Casas}, {Federici},
  {Fedorets}, {Fernique}, {Figueras}, {Filippi}, {Findeisen}, {Fonti},
  {Fraile}, {Fraser}, {Fr{\'e}zouls}, {Gai}, {Galleti}, {Garabato},
  {Garc{\'\i}a-Sedano}, {Garofalo}, {Garralda}, {Gavel}, {Gavras}, {Gerssen},
  {Geyer}, {Giacobbe}, {Gilmore}, {Girona}, {Giuffrida}, {Glass}, {Gomes},
  {Granvik}, {Gueguen}, {Guerrier}, {Guiraud}, {Guti{\'e}rrez-S{\'a}nchez},
  {Haigron}, {Hatzidimitriou}, {Hauser}, {Haywood}, {Heiter}, {Helmi}, {Heu},
  {Hilger}, {Hobbs}, {Hofmann}, {Holland}, {Huckle}, {Hypki}, {Icardi},
  {Jan{\ss}en}, {Jevardat de Fombelle}, {Jonker}, {Juh{\'a}sz}, {Julbe},
  {Karampelas}, {Kewley}, {Klar}, {Kochoska}, {Kohley}, {Kolenberg},
  {Kontizas}, {Kontizas}, {Koposov}, {Kordopatis}, {Kostrzewa-Rutkowska},
  {Koubsky}, {Lambert}, {Lanza}, {Lasne}, {Lavigne}, {Le Fustec}, {Le
  Poncin-Lafitte}, {Lebreton}, {Leccia}, {Leclerc}, {Lecoeur-Taibi},
  {Lenhardt}, {Leroux}, {Liao}, {Licata}, {Lindstr{\o}m}, {Lister}, {Livanou},
  {Lobel}, {L{\'o}pez}, {Managau}, {Mann}, {Mantelet}, {Marchal}, {Marchant},
  {Marconi}, {Marinoni}, {Marschalk{\'o}}, {Marshall}, {Martino}, {Marton},
  {Mary}, {Massari}, {Matijevi{\v{c}}}, {Mazeh}, {McMillan}, {Messina},
  {Michalik}, {Millar}, {Molina}, {Molinaro}, {Moln{\'a}r}, {Montegriffo},
  {Mor}, {Morbidelli}, {Morel}, {Morris}, {Mulone}, {Muraveva}, {Musella},
  {Nelemans}, {Nicastro}, {Noval}, {O'Mullane}, {Ord{\'e}novic},
  {Ord{\'o}{\~n}ez-Blanco}, {Osborne}, {Pagani}, {Pagano}, {Pailler},
  {Palacin}, {Palaversa}, {Panahi}, {Pawlak}, {Piersimoni}, {Pineau}, {Plachy},
  {Plum}, {Poggio}, {Poujoulet}, {Pr{\v{s}}a}, {Pulone}, {Racero}, {Ragaini},
  {Rambaux}, {Ramos-Lerate}, {Regibo}, {Reyl{\'e}}, {Riclet}, {Ripepi}, {Riva},
  {Rivard}, {Rixon}, {Roegiers}, {Roelens}, {Romero-G{\'o}mez}, {Rowell},
  {Royer}, {Ruiz-Dern}, {Sadowski}, {Sagrist{\`a} Sell{\'e}s}, {Sahlmann},
  {Salgado}, {Salguero}, {Sanna}, {Santana-Ros}, {Sarasso}, {Savietto},
  {Schultheis}, {Sciacca}, {Segol}, {Segovia}, {S{\'e}gransan}, {Shih},
  {Siltala}, {Silva}, {Smart}, {Smith}, {Solano}, {Solitro}, {Sordo}, {Soria
  Nieto}, {Souchay}, {Spagna}, {Spoto}, {Stampa}, {Steele},
  {Steidelm{\"u}ller}, {Stephenson}, {Stoev}, {Suess}, {Surdej}, {Szabados},
  {Szegedi-Elek}, {Tapiador}, {Taris}, {Tauran}, {Taylor}, {Teixeira},
  {Terrett}, {Teyssandier}, {Thuillot}, {Titarenko}, {Torra Clotet}, {Turon},
  {Ulla}, {Utrilla}, {Uzzi}, {Vaillant}, {Valentini}, {Valette}, {van Elteren},
  {Van Hemelryck}, {van Leeuwen}, {Vaschetto}, {Vecchiato}, {Veljanoski},
  {Viala}, {Vicente}, {Vogt}, {von Essen}, {Voss}, {Votruba}, {Voutsinas},
  {Walmsley}, {Weiler}, {Wertz}, {Wevers}, {Wyrzykowski}, {Yoldas},
  {{\v{Z}}erjal}, {Ziaeepour}, {Zorec}, {Zschocke}, {Zucker}, {Zurbach}, \&
  {Zwitter}}]{brown2018}
{Gaia Collaboration}, {Brown}, A.~G.~A., {Vallenari}, A., {et~al.} 2018, \aap,
  616, A1

\bibitem[{{Gil-Pons} {et~al.}(2022){Gil-Pons}, {Doherty}, {Campbell}, \&
  {Guti{\'e}rrez}}]{gilpons2022}
{Gil-Pons}, P., {Doherty}, C.~L., {Campbell}, S.~W., \& {Guti{\'e}rrez}, J.
  2022, \aap, 668, A100

\bibitem[{{Gil-Pons} {et~al.}(2021){Gil-Pons}, {Doherty}, {Guti{\'e}rrez},
  {Campbell}, {Siess}, \& {Lattanzio}}]{gilpons2021}
{Gil-Pons}, P., {Doherty}, C.~L., {Guti{\'e}rrez}, J., {et~al.} 2021, \aap,
  645, A10

\bibitem[{{Gil-Pons} {et~al.}(2018){Gil-Pons}, {Doherty}, {Guti{\'e}rrez},
  {Siess}, {Campbell}, {Lau}, \& {Lattanzio}}]{gilpons2018}
{Gil-Pons}, P., {Doherty}, C.~L., {Guti{\'e}rrez}, J.~L., {et~al.} 2018, \pasa,
  35

\bibitem[{{Grevesse} {et~al.}(1996){Grevesse}, {Noels}, \&
  {Sauval}}]{grevesse1996}
{Grevesse}, N., {Noels}, A., \& {Sauval}, A.~J. 1996, in Astronomical Society
  of the Pacific Conference Series, Vol.~99, Cosmic Abundances, ed. S.~S.
  {Holt} \& G.~{Sonneborn}, 117

\bibitem[{{Hansen} {et~al.}(2016){Hansen}, {Andersen}, {Nordstr{\"o}m},
  {Beers}, {Placco}, {Yoon}, \& {Buchhave}}]{Hansen2016b}
{Hansen}, T.~T., {Andersen}, J., {Nordstr{\"o}m}, B., {et~al.} 2016, \aap, 588,
  A3

\bibitem[{{Heil} {et~al.}(2008){Heil}, {Detwiler}, {Azuma}, {Couture}, {Daly},
  {G{\"o}rres}, {K{\"a}ppeler}, {Reifarth}, {Tischhauser}, {Ugalde}, \&
  {Wiescher}}]{heil2008}
{Heil}, M., {Detwiler}, R., {Azuma}, R.~E., {et~al.} 2008, \prc, 78, 025803

\bibitem[{{Herwig}(2004{\natexlab{a}})}]{herwig2004}
{Herwig}, F. 2004{\natexlab{a}}, \apj, 605, 425

\bibitem[{{Herwig}(2004{\natexlab{b}})}]{her04b}
{Herwig}, F. 2004{\natexlab{b}}, \apjs, 155, 651

\bibitem[{{Herwig} {et~al.}(1997){Herwig}, {Bloecker}, {Schoenberner}, \& {El
  Eid}}]{Herwig1997}
{Herwig}, F., {Bloecker}, T., {Schoenberner}, D., \& {El Eid}, M. 1997, \aap,
  324, L81

\bibitem[{{Iliadis} {et~al.}(2010){Iliadis}, {Longland}, {Champagne}, {Coc}, \&
  {Fitzgerald}}]{iliadis2010}
{Iliadis}, C., {Longland}, R., {Champagne}, A.~E., {Coc}, A., \& {Fitzgerald},
  R. 2010, \nphysa, 841, 31

\bibitem[{{Ishigaki} {et~al.}(2014){Ishigaki}, {Tominaga}, {Kobayashi}, \&
  {Nomoto}}]{ishigaki2014}
{Ishigaki}, M.~N., {Tominaga}, N., {Kobayashi}, C., \& {Nomoto}, K. 2014,
  \apjl, 792, L32

\bibitem[{Iwamoto {et~al.}(2005)Iwamoto, Umeda, Tominaga, Nomoto, \&
  Maeda}]{iwamoto2005}
Iwamoto, N., Umeda, H., Tominaga, N., Nomoto, K., \& Maeda, K. 2005, Science,
  309, 451

\bibitem[{{Izzard} {et~al.}(2009){Izzard}, {Glebbeek}, {Stancliffe}, \&
  {Pols}}]{izzard2009}
{Izzard}, R.~G., {Glebbeek}, E., {Stancliffe}, R.~J., \& {Pols}, O.~R. 2009,
  \aap, 508, 1359

\bibitem[{Jeena {et~al.}(2023)Jeena, Banerjee, Chiaki, \& Heger}]{jeena2023}
Jeena, S.~K., Banerjee, P., Chiaki, G., \& Heger, A. 2023, Monthly Notices of
  the Royal Astronomical Society, 526, 4467

\bibitem[{{Karakas} \& {Lattanzio}(2007)}]{karakas2007}
{Karakas}, A. \& {Lattanzio}, J.~C. 2007, \pasa, 24, 103

\bibitem[{{Karakas}(2010)}]{karakas2010}
{Karakas}, A.~I. 2010, \mnras, 403, 1413

\bibitem[{{Karakas} \& {Lattanzio}(2014)}]{karakas2014}
{Karakas}, A.~I. \& {Lattanzio}, J.~C. 2014, \pasa, 31, e030

\bibitem[{{Kobayashi} {et~al.}(2020){Kobayashi}, {Karakas}, \&
  {Lugaro}}]{kobayashi2020}
{Kobayashi}, C., {Karakas}, A.~I., \& {Lugaro}, M. 2020, \apj, 900, 179

\bibitem[{{Korn} {et~al.}(2009){Korn}, {Richard}, {Mashonkina}, {Bessell},
  {Frebel}, \& {Aoki}}]{korn2009}
{Korn}, A.~J., {Richard}, O., {Mashonkina}, L., {et~al.} 2009, \apj, 698, 410

\bibitem[{{Lattanzio} {et~al.}(1996){Lattanzio}, {Frost}, {Cannon}, \&
  {Wood}}]{Lattanzio1996}
{Lattanzio}, J., {Frost}, C., {Cannon}, R., \& {Wood}, P.~R. 1996, \memsai, 67,
  729

\bibitem[{{Lattanzio}(1986)}]{lat86}
{Lattanzio}, J.~C. 1986, \apj, 311, 708

\bibitem[{{Lederer} \& {Aringer}(2009)}]{lederer2009}
{Lederer}, M.~T. \& {Aringer}, B. 2009, \aap, 494, 403

\bibitem[{{Lucatello} {et~al.}(2005){Lucatello}, {Tsangarides}, {Beers},
  {Carretta}, {Gratton}, \& {Ryan}}]{luc05}
{Lucatello}, S., {Tsangarides}, S., {Beers}, T.~C., {et~al.} 2005, \apj, 625,
  825

\bibitem[{{Lugaro} {et~al.}(2012){Lugaro}, {Karakas}, {Stancliffe}, \&
  {Rijs}}]{lugaro2012}
{Lugaro}, M., {Karakas}, A.~I., {Stancliffe}, R.~J., \& {Rijs}, C. 2012, \apj,
  747, 2

\bibitem[{{Maeda} \& {Nomoto}(2003)}]{maeda2003}
{Maeda}, K. \& {Nomoto}, K. 2003, \apj, 598, 1163

\bibitem[{{Maeder} \& {Meynet}(2015)}]{maeder2015}
{Maeder}, A. \& {Meynet}, G. 2015, \aap, 580, A32

\bibitem[{{Marigo} \& {Aringer}(2009)}]{marigo2009}
{Marigo}, P. \& {Aringer}, B. 2009, \aap, 508, 1539

\bibitem[{{Mashonkina} {et~al.}(2017){Mashonkina}, {Sitnova}, \&
  {Belyaev}}]{mashonkina2017}
{Mashonkina}, L., {Sitnova}, T., \& {Belyaev}, A.~K. 2017, \aap, 605, A53

\bibitem[{{Masseron} {et~al.}(2020){Masseron}, {Garc{\'\i}a-Hern{\'a}ndez},
  {Santove{\~n}a}, {Manchado}, {Zamora}, {Manteiga}, \&
  {Dafonte}}]{masseron2020}
{Masseron}, T., {Garc{\'\i}a-Hern{\'a}ndez}, D.~A., {Santove{\~n}a}, R.,
  {et~al.} 2020, Nature Communications, 11, 3759

\bibitem[{{Meynet} {et~al.}(2006){Meynet}, {Ekstr{\"o}m}, \& {Maeder}}]{mey06}
{Meynet}, G., {Ekstr{\"o}m}, S., \& {Maeder}, A. 2006, \aap, 447, 623

\bibitem[{{Molaro} {et~al.}(2023){Molaro}, {Aguado}, {Caffau}, {Allende
  Prieto}, {Bonifacio}, {Gonz{\'a}lez Hern{\'a}ndez}, {Rebolo}, {Zapatero
  Osorio}, {Cristiani}, {Pepe}, {Santos}, {Alibert}, {Cupani}, {Di
  Marcantonio}, {D'Odorico}, {Lovis}, {Martins}, {Milakovi{\'c}}, {Murphy},
  {Nunes}, {Schmidt}, {Sousa}, {Sozzetti}, \& {Su{\'a}rez
  Mascare{\~n}o}}]{molaro2023}
{Molaro}, P., {Aguado}, D.~S., {Caffau}, E., {et~al.} 2023, \aap, 679, A72

\bibitem[{{Nandal} {et~al.}(2024){Nandal}, {Sibony}, \&
  {Tsiatsiou}}]{Nandal2024}
{Nandal}, D., {Sibony}, Y., \& {Tsiatsiou}, S. 2024, \aap, 688, A142

\bibitem[{{Pols} {et~al.}(2012){Pols}, {Izzard}, {Stancliffe}, \&
  {Glebbeek}}]{pols2012}
{Pols}, O.~R., {Izzard}, R.~G., {Stancliffe}, R.~J., \& {Glebbeek}, E. 2012,
  \aap, 547, A76

\bibitem[{{Ritter} {et~al.}(2018){Ritter}, {Herwig}, {Jones}, {Pignatari},
  {Fryer}, \& {Hirschi}}]{ritter2018}
{Ritter}, C., {Herwig}, F., {Jones}, S., {et~al.} 2018, \mnras, 480, 538

\bibitem[{{Ritter} {et~al.}(2012){Ritter}, {Safranek-Shrader}, {Gnat},
  {Milosavljevi{\'c}}, \& {Bromm}}]{ritter12}
{Ritter}, J.~S., {Safranek-Shrader}, C., {Gnat}, O., {Milosavljevi{\'c}}, M.,
  \& {Bromm}, V. 2012, \apj, 761, 56

\bibitem[{{Roberti} {et~al.}(2024){Roberti}, {Limongi}, \&
  {Chieffi}}]{roberti2024}
{Roberti}, L., {Limongi}, M., \& {Chieffi}, A. 2024, \apjs, 270, 28

\bibitem[{{Saito} {et~al.}(2009){Saito}, {Takada-Hidai}, {Honda}, \&
  {Takeda}}]{Saito2009}
{Saito}, Y.-J., {Takada-Hidai}, M., {Honda}, S., \& {Takeda}, Y. 2009, \pasj,
  61, 549

\bibitem[{{Sharda} {et~al.}(2021){Sharda}, {Federrath}, {Krumholz}, \&
  {Schleicher}}]{Sharda2021}
{Sharda}, P., {Federrath}, C., {Krumholz}, M.~R., \& {Schleicher}, D. R.~G.
  2021, \mnras, 503, 2014

\bibitem[{{Straniero} {et~al.}(2014){Straniero}, {Cristallo}, \&
  {Piersanti}}]{straniero2014}
{Straniero}, O., {Cristallo}, S., \& {Piersanti}, L. 2014, \apj, 785, 77

\bibitem[{{Straniero} {et~al.}(1995){Straniero}, {Gallino}, {Busso}, {Chiefei},
  {Raiteri}, {Limongi}, \& {Salaris}}]{straniero1995}
{Straniero}, O., {Gallino}, R., {Busso}, M., {et~al.} 1995, \apjl, 440, L85

\bibitem[{{Takahashi} {et~al.}(2014){Takahashi}, {Umeda}, \&
  {Yoshida}}]{takahashi2014}
{Takahashi}, K., {Umeda}, H., \& {Yoshida}, T. 2014, \apj, 794, 40

\bibitem[{{Tominaga} {et~al.}(2014){Tominaga}, {Iwamoto}, \&
  {Nomoto}}]{tominaga2014}
{Tominaga}, N., {Iwamoto}, N., \& {Nomoto}, K. 2014, \apj, 785, 98

\bibitem[{{Tominaga} {et~al.}(2007){Tominaga}, {Umeda}, \&
  {Nomoto}}]{tominaga2007}
{Tominaga}, N., {Umeda}, H., \& {Nomoto}, K. 2007, \apj, 660, 516

\bibitem[{{Tsiatsiou} {et~al.}(2024){Tsiatsiou}, {Sibony}, {Nandal},
  {Sciarini}, {Hirai}, {Ekstr{\"o}m}, {Farrell}, {Murphy}, {Choplin},
  {Hirschi}, {Chiappini}, {Liu}, {Bromm}, {Groh}, \& {Meynet}}]{Tsiatsiou2024}
{Tsiatsiou}, S., {Sibony}, Y., {Nandal}, D., {et~al.} 2024, \aap, 687, A307

\bibitem[{{Umeda} \& {Nomoto}(2002)}]{Umeda2002}
{Umeda}, H. \& {Nomoto}, K. 2002, \apj, 565, 385

\bibitem[{{Umeda} \& {Nomoto}(2003)}]{umeda2003}
{Umeda}, H. \& {Nomoto}, K. 2003, \nat, 422, 871

\bibitem[{{Ventura} {et~al.}(2001){Ventura}, {D'Antona}, {Mazzitelli}, \&
  {Gratton}}]{ventura2001}
{Ventura}, P., {D'Antona}, F., {Mazzitelli}, I., \& {Gratton}, R. 2001, \apjl,
  550, L65

\bibitem[{{Yong} {et~al.}(2013){Yong}, {Norris}, {Bessell}, {Christlieb},
  {Asplund}, {Beers}, {Barklem}, {Frebel}, \& {Ryan}}]{yong13}
{Yong}, D., {Norris}, J.~E., {Bessell}, M.~S., {et~al.} 2013, \apj, 762, 27

\bibitem[{{Yoon} {et~al.}(2016){Yoon}, {Beers}, {Placco}, {Rasmussen},
  {Carollo}, {He}, {Hansen}, {Roederer}, \& {Zeanah}}]{Yoon2016}
{Yoon}, J., {Beers}, T.~C., {Placco}, V.~M., {et~al.} 2016, \apj, 833, 20

\end{thebibliography}
%


\newpage
\begin{appendix} 
\label{sec:appA}

\appendix

\section{Details of comparisons between models and observations}\label{app:details}

\subsection{Dilution}

According to our proposed scenario, the ejecta from our intermediate-mass stars are mixed with the natal cloud material.
 
We define the dilution factor $f_{dil}$:
\begin{equation}
    f_{dil}=\frac{M_{ej}}{M_{ej}+M_{cl}},
    \label{eq:f}
\end{equation}
\noindent where $M_{ej}$ is the mass ejected by one of our model stars, and $M_{cl}$ is the mass of the natal cloud. 
Taking $M_{i,ej}$ and $M_{i,cl}$, respectively, as the ejected and cloud masses of species `i', the mass fraction of each species is:
\begin{equation}
    X_{i,\star}=\frac{M_{i,ej}+M_{i,cl}}{M_{ej} + M_{cl}}.
\end{equation}

Thus, $X_{i,\star}$ can be expressed in terms of the ejecta abundances $X_{i,ej}$ and cloud abundances $X_{i,cl}$:
\begin{equation}
    X_{i,\star}=f_{dil} X_{i,ej} + (1-f_{dil})X_{i,cl}
\end{equation}

After converting from mass to number abundances ($Y_{i,ej}$ and $Y_{i,cl}$) our theoretical abundance of each species in the standard notation becomes:
\begin{equation}
    \rm{[X_i/Fe]}=\log_{10}\left(\frac{Y_i}{\rm{Y_{Fe}}} \right)_\star - \log_{10}\left(\frac{Y_i}{Y_{Fe}} \right)_\odot. 
\end{equation}

The results of the former theoretical yield abundances for some selected elements, for different initial masses ($M_{ini}$), and for different masses of the partially mixing zone $\Delta m_{PMZ}$ are shown in Table \ref{tab:yields}.

\subsection{Abundance pattern comparisons: $\chi_N^2$ test}

We calculated $\chi_N^2$ as in \citealt{bisterzo2011}, to assess and compare the quality of our fits and others existing in the literature.

\begin{equation}
    \chi_N^2=\sum_{i=1}^N \frac{\Bigl(\rm{[X_i/Fe]} - \rm{[X_i/Fe]_{obs}}\Bigr)^2}{\epsilon_i^2}
\end{equation}

\noindent where [$X_i/Fe$] have been described above, $[X_i/Fe]_{obs}$ are the analogous observed abundances, $\epsilon$ is the observational error corresponding to $2\sigma$, and $N$ is the number of considered elements in the observed data (with error bars).

Table \ref{tab:chi2} summarises our results for the $\chi^2$ test for different cases. Case A considers all elements detected (i.e. excluding those with upper limits). Case B considers the same species as Case A but excludes the single element that provides the worst fit between theoretical and observed abundances. This case allows us to disregard the effects of possible outliers, which are interpreted as those species whose abundances may be particularly problematic to determine observationally or are especially sensitive to theoretical input physics and uncertainties. Case C further relaxes the statistical comparison, excluding the two worst-fitting elements. Moreover, this case allows us to consider the comparison between the observed data and the supernova models by \citealt{ezzeddine2019} and \citealt{jeena2023}, who did not account for the presence of Sr and Ba in their fits.

The results of the $\chi^2$-test, shown in Table \ref{tab:chi2} imply the following:

\begin{itemize}

    \item[i)] When all the elements are considered (Case A), we obtain the best probability distribution (or confidence level) value for the intermediate-mass models. Our best-fitting model (3~\msun{} with a $m_{PMZ}=10^{-3}$~\msun) yields $P_n(\chi^2)=52\%$, whereas the SN models from \citealt{ezzeddine2019} and \citealt{jeena2023}, unable to produce heavy elements, yield $P_n(\chi^2)$ below 10$\%$. 

    \item[ii)] When we discard the single worst-fitting element (Case B), the $P_n(\chi^2)$ value for our 3~\msun{} with a $m_{PMZ}=10^{-3}$~\msun{} increases up to 81$\%$. Oxygen is the element that affects most significantly the quality of the fits of all the 3~\msun{} models. As reported in Section \ref{sec:compare_ours}, oxygen is a difficult species to measure in the case of HE~1327. On the other hand, our models naturally suffer from input physics uncertainties that affect theoretical abundances. A more efficient TDU, that could be achieved introducing diffusive overshooting at the base of the pulse-driven convective zone \citep{herwig2004}, or a number of TDU episodes higher than the ones we computed (less efficient stellar winds could cause that) would lead to higher theoretical oxygen abundances, together with mildly higher heavy-element abundances. Both trends would improve even more the fit to observed values. The fit of supernova models when disregarding Ba increases to $P_n(\chi^2)=28\%$ and $51\%$ for the models by \citealt{ezzeddine2019} and \citealt{jeena2023}, respectively.

    \item [iii)] Case C (when we disregard the \textit{two} worst-fitting species) recovers very high confidence levels for the supernova models. However, our best-fitting model performs considerably better than the SN fit in \citealt{ezzeddine2019} ($P_n(\chi^2)=94\%$ versus $86\%$), and very close to the fit in \citealt{jeena2023}, with $P_n(\chi^2)=97\%$.

  \end{itemize}

To further assess our comparison analysis, we calculated modified versions of the $\chi^2$ test. First, we considered including species for which observations have determined the upper limits. If theoretical abundances remained below these limits (as was the case for all the considered models), no term was added to calculate the $\chi^2$ value, but the number of degrees of freedom increased to account for species with upper limits. Consequently, the confidence level values improved considerably. For Case A, the intermediate-mass best fit model yielded $P_n(\chi^2)=87\%$, and the best SN fit \citep{jeena2023} yielded $P_n(\chi^2)=35\%$. For Case B, $P_n(\chi^2)=98\%$ and $76\%$ for the best-fit intermediate-mass and supernova models, respectively. Finally, all models yielded $P_n(\chi^2)\geq 97\%$ for Case C.

We also calculated the modified $\chi^2$ reported in \citealt{jeena2023}, who added an offset term to minimise $\chi^2$, when all elements with upper limits were ignored but still retained all the species (both with error bars and upper limits) in the count to calculate their $\chi^2$-values. This naturally reduced all $\chi_n^2$ to very low values ($\lesssim 1$), and all $P_n(\chi^2)\simeq100\%$.  
Therefore, we found that this method was not suitable for discriminating between the different models we considered.

\begin{table*}
\begin{center}
\begingroup
\setlength{\tabcolsep}{4pt} 
\begin{tabular}{ lccr ccr ccr}
            \noalign{\smallskip}
            \hline
            \noalign{\smallskip}
 & \multicolumn{3}{c}{Case A} &  \multicolumn{3}{c}{Case B} & \multicolumn{3}{c}{Case C}\\
 Model & \multicolumn{3}{c}{{\small (All elements)}}  & \multicolumn{3}{c}{{{\small (All minus 1 elem.)}}} & \multicolumn{3}{c}{{{\small (All minus 2 elem.)}}}\\
   & N & $\chi_n^2$ & $P(\chi_n^2)$ & N &  $\chi_n^2$ & $P(\chi_n^2)$ & N & $\chi_n^2$ & $P(\chi_n^2)$  \\
            \noalign{\smallskip}
            \hline
            \noalign{\smallskip}
\noalign{\smallskip}
\multicolumn{10}{l}{$2\sigma$, no upper limits}\\
\noalign{\smallskip}
3.0 \msun{}, no OV, {\small $\Delta m_{PMZ}=10^{-3}$\msun}           & 14  & 13.1 & 52$\%$  & 13 (-O) & 8.4 & 81$\%$  & 11 (-O, Mg) & 5.5 & 94$\%$ \\
3.0 \msun{}, no OV, {\small $\Delta m_{PMZ}=5\times 10^{-4}$\msun}   & 14  & 15.1 & 37$\%$  & 13 (-O) & 10.1 & 68$\%$  & 12 (-O, Mg) & 8.6 & 74$\%$ \\
3.2 \msun{}, OV, {\small $\Delta m_{PMZ}=6\times 10^{-4}$\msun}   & 14  & 15.2 & 36$\%$  & 13 (-O) & 10.1 & 69$\%$  & 12 (-O, Mg) & 7.0 & 86$\%$ \\
3.5 \msun{}, no OV, {\small $\Delta m_{PMZ}=5\times 10^{-4}$\msun}   & 14  & 15.5 & 35$\%$  & 12 (-O) & 10.2 & 68$\%$  & 11 (-O, Mg) & 7.5 & 82$\%$ \\

\noalign{\smallskip}
Aspherical SN (Ezz19) &   14  & 25.0 & 3$\%$ & 13 (-Ba) & 17 & 28$\%$ & 12 (-Sr, Ba) & 7.1 & 85$\%$ \\
Rot QCH SN (Jen23) &   13  & 21.8 & 6$\%$ & 12 (-Ba) & 11.3 & 51$\%$ &  11 (-Sr, Ba) & 4.0 & 97$\%$\\

\hline
\noalign{\smallskip}
\noalign{\smallskip}
\multicolumn{10}{l}{$2\sigma$, including upper limits}\\
\noalign{\smallskip}
3.0 \msun{}, no OV, {\small $\Delta m_{PMZ}=10^{-3}$\msun}           & 20  & 13.1 & 87$\%$  & 12 (-O) & 8.4 & 98$\%$  & 11 (-O, Mg) & 5.5 & >99$\%$ \\
3.0 \msun{}, no OV, {\small $\Delta m_{PMZ}=5\times 10^{-4}$\msun}   & 20  & 15.1 & 77$\%$  & 12 (-O) & 10.1 & 95$\%$  & 11 (-O, Mg) & 8.6 & 97$\%$ \\
3.2 \msun{}, OV, {\small $\Delta m_{PMZ}=6\times 10^{-4}$\msun}   & 20  & 15.2 & 81$\%$  & 12 (-O) & 10.1 & 97$\%$  & 11 (-O, Mg) & 7.0 & 99$\%$ \\
3.5 \msun{}, no OV, {\small $\Delta m_{PMZ}=5\times 10^{-4}$\msun}   & 20  & 15.5 & 75$\%$  & 12 (-O) & 10.2 & 95$\%$  & 11 (-O, Mg) & 7.5 & 98$\%$ \\

\noalign{\smallskip}
Aspherical SN (Ezz19) &   21  & 25.0 & 20$\%$ & 20 (-Ba) & 14.4 & 76$\%$ & 19 (-Sr, Ba) & 7.1 & 99$\%$ \\
Rot QCH SN (Jen23) &   20  & 21.8 & 35$\%$ & 19 (-Ba) & 14.5 & 75$\%$ &  18 (-Sr, Ba) & 4.0 & >99$\%$\\
\hline

\end{tabular}

\caption{Results for the $\chi_N^2$ performed for different models. $N$ is the number of degrees of freedom, $\chi_N^2$ is the $\chi-$squared value of the comparison, $P(\chi_N^2)$ is the probability distribution (a proxy for confidence level) obtained from the standard $\chi^2$ tables \citep{standardmat1991}. For case A, we considered all the available observed values. For cases $B$ and $C$, we disregarded, respectively, the first and the first two elements that yielded the worst (highest) $q_i=\frac{([X_i/Fe]-[X_i/Fe]_{obs})^2}{\epsilon_i^2}$ values (see text for details). 
The citation keys for the different models are the same as in Figure \ref{fig:Comp_other}. 
}
\label{tab:chi2}
\endgroup  
\end{center}
\end{table*}

\subsection{Summary of $\chi^2$ comparisons}

In conclusion, we found that the intermediate-mass models fit the observed abundances of HE~1327 considerably better than the supernova models when all the observed species (or all except the worst-fitting species) are considered. This occurs independently of whether we consider elements for which only upper limits were determined. When we exclude the two worst-fitting elements, the best-fitting supernova model performs just slightly better than the best-fitting intermediate-mass model.

\begin{figure*}[h]
    \centering
\includegraphics[width=\linewidth]{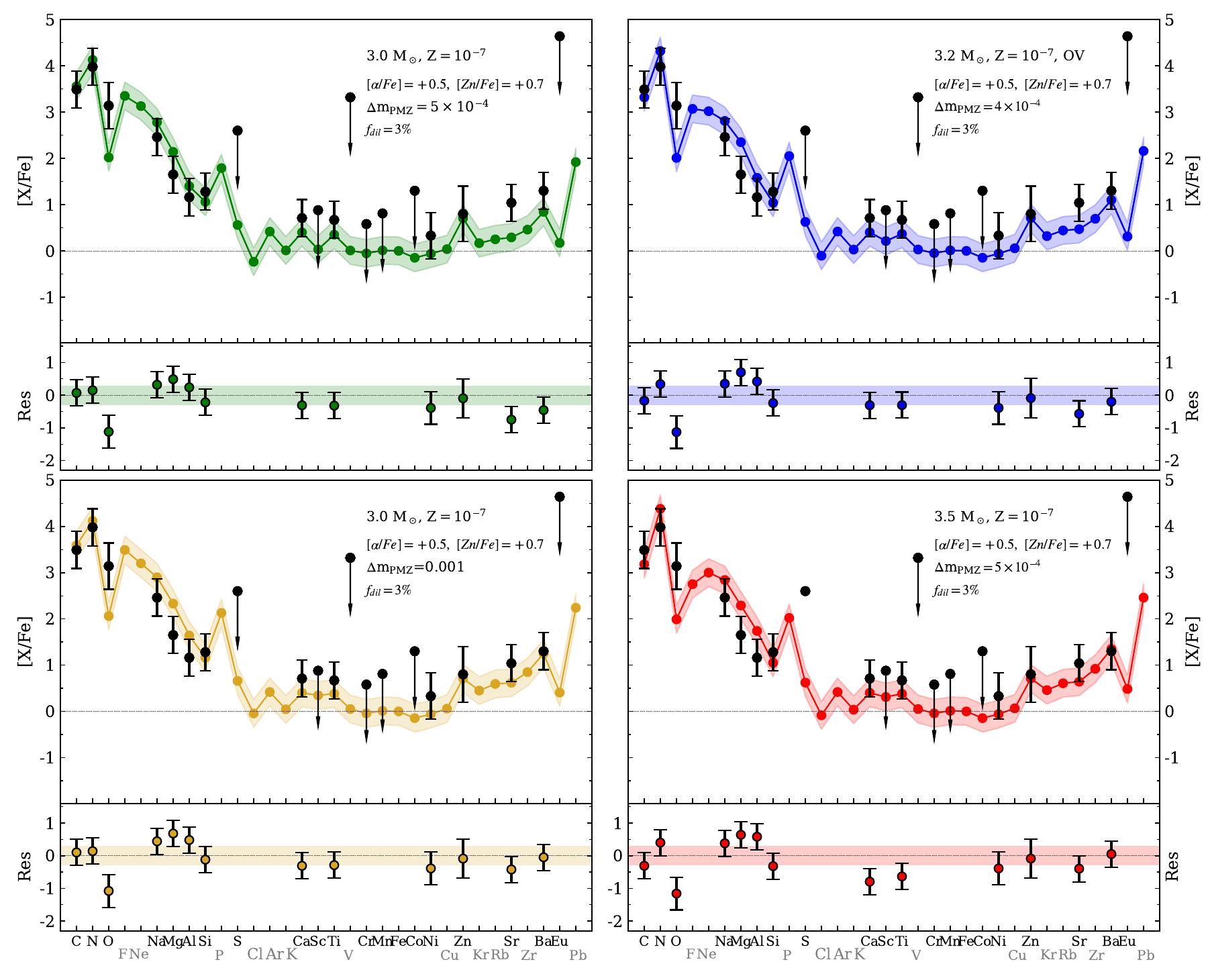}
      \caption{Theoretical yield abundances (post-dilution) are plotted together with the observed abundances of HE~1327, as in Fig.~\ref{fig:Comp_obs}. See Table~\ref{tab:yields} for numerical values. Model parameters are given in each panel (see also Section~\ref{sec:models} and Table~\ref{tab:mod}). The residuals and shaded areas have the same meaning as in Figure \ref{fig:Comp_obs}. The $\chi^2$ values are given in Table~\ref{tab:chi2}.}
         \label{fig:Comp_ours_obs}
\end{figure*}

We present the abundance patterns visually, comparing them to observations for our models in Figure~\ref{fig:Comp_ours_obs}. For the newest supernova and low-mass models from the literature, we provide a comparison in Figure~\ref{fig:Comp_other}.

\subsection{Model yield uncertainties}
\label{appx:uncerts}
In Figure~\ref{fig:Comp_obs}, we attempt to highlight the uncertainties in model yields by including some shading around our values. These uncertainties primarily stem from our limited understanding of internal mixing processes, the location of convective boundaries, mass-loss rates, and, in some cases, nuclear reaction rates. Quantifying these uncertainties is complicated, so we opted to consider the range of yield variations from models by different authors in the existing literature. For instance, \cite{ritter2018} compared results from \cite{karakas2010} and \cite{straniero2014} for of 2 \msun models of $Z=10^{-4}$. The yield differences in dex are $\Delta_{KS}(C)=|Log(Y_{K10}(C) - Log(Y_{S14}(C))|=0.36$. Analogously, for N and O, $\Delta_{KS}(N)=0.26$ and $\Delta_{KS}(O)=0.30$. Comparisons between the yields from \cite{ritter2018} and \cite{karakas2010} yielded similar values (somewhat lower for C), or considerably higher ($\Delta_{KR}(O)\simeq 1$). Comparing published yields for 3 \msun~ of $Z=10^{-4}$ models yields $\Delta_{KS}(C)=0.30$, $\Delta_{KS}(N)=0.13$ and $\Delta_{KS}(O)=0.07$, and     $\Delta_{KR}(C)=0.42$, $\Delta_{KR}(N)=0.41$ and $\Delta_{KR}(O)=0.72$. Considering that uncertainties increase even more for lower metallicity models, we take a $\Delta \pm 0.3$ dex as a conservative value for our analysis. As a reference, \cite{gilpons2021} analysed the effects of different standard wind prescriptions on stellar yields and found variations that, in all cases, were higher than 0.4 dex. For more on model uncertainties we refer the interested reader to \cite{karakas2014} for a general overview and to \cite{gilpons2018} for the specific case of extremely low-metallicity stars.

\newpage
\section{Model abundances table}

In Table~\ref{tab:yields} we list the elemental yields for our grid of models, after dilution with the primitive cloud composition (see Section~\ref{sec:models}).

\begin{table*}[h]
\begin{center}
    \begin{tabular}{cr@{\hskip 0.5cm}rrr@{\hskip 0.9cm}r@{\hskip 1.0cm}r@{\hskip 1.0cm}r}
            \hline
            \noalign{\smallskip}
   &  & & &  & [X/Fe] &   &   \\
  \noalign{\smallskip}
  \cline{3-8}
            \noalign{\smallskip}
    &  & $M_{ini}$/\msun & & 3.0  & 3.0 & 3.2  & 3.5  \\
  \noalign{\smallskip}
 & & OV & & No & No & Yes & No\\
            \noalign{\smallskip}
 Element          &  & $\Delta m_{PMZ}$/\msun    & $0$ & 
                $5\times 10^{-4}$ & 
                $10^{-3}$ & 
                $6\times 10^{-4}$ & 
                $5\times 10^{-4}$ \\
                            \noalign{\smallskip}
                            \hline
            \noalign{\smallskip}
                  C    & &  &   3.79 &   3.56 &   3.66 &   3.32 &   3.19 \\
                  N    & &  &   4.26 &   4.13 &   4.19 &   4.32 &   4.37 \\
                  O    & &  &   1.97 &   2.02 &   2.13 &   2.01 &   1.99 \\
                  F    & &  &   3.14 &   3.35 &   3.56 &   3.07 &   2.75 \\
                 Ne    & &  &   2.70 &   3.13 &   3.27 &   3.02 &   3.00 \\
                 Na    & &  &   2.36 &   2.78 &   2.96 &   2.81 &   2.85 \\
                 Mg    & &  &   1.83 &   2.14 &   2.40 &   2.35 &   2.29 \\
                 Al    & &  &   1.07 &   1.40 &   1.71 &   1.58 &   1.75 \\
                 Si    & &  &   0.74 &   1.06 &   1.22 &   1.04 &   1.05 \\
                  P    & &  &   0.84 &   1.79 &   2.19 &   2.05 &   2.01 \\
                  S    & &  &   0.48 &   0.56 &   0.68 &   0.63 &   0.62 \\
                 Cl    & &  &  -0.31 &  -0.24 &  -0.01 &  -0.10 &  -0.08 \\
                 Ar    & &  &   0.42 &   0.42 &   0.42 &   0.42 &   0.42 \\
                  K    & &  &   0.00 &   0.01 &   0.05 &   0.03 &   0.04 \\
                 Ca    & &  &   0.40 &   0.40 &   0.40 &   0.40 &   0.40 \\
                 Sc    & &  &  -0.13 &   0.03 &   0.39 &   0.21 &   0.32 \\
                 Ti    & &  &   0.33 &   0.35 &   0.39 &   0.37 &   0.36 \\
                  V    & &  &  -0.01 &   0.01 &   0.06 &   0.03 &   0.04 \\
                 Cr    & &  &  -0.06 &  -0.05 &  -0.04 &  -0.05 &  -0.04 \\
                 Mn    & &  &   0.01 &   0.01 &   0.01 &   0.01 &   0.01 \\
                 Fe    & &  &   0.00 &   0.00 &   0.00 &   0.00 &   0.00 \\
                 Co    & &  &  -0.15 &  -0.15 &  -0.14 &  -0.15 &  -0.14 \\
                 Ni    & &  &  -0.06 &  -0.06 &  -0.06 &  -0.06 &  -0.06 \\
                 Cu    & &  &   0.04 &   0.04 &   0.07 &   0.06 &   0.07 \\
                 Zn    & &  &   0.70 &   0.70 &   0.71 &   0.71 &   0.70 \\
                 Kr    & &  &   0.02 &   0.17 &   0.50 &   0.32 &   0.49 \\
                 Rb    & &  &   0.05 &   0.25 &   0.65 &   0.44 &   0.62 \\
                 Sr    & &  &   0.03 &   0.29 &   0.67 &   0.47 &   0.64 \\
                 Zr    & &  &  -0.03 &   0.46 &   0.91 &   0.69 &   0.93 \\
                 Ba    & &  &  -0.02 &   0.84 &   1.32 &   1.10 &   1.36 \\
                 Eu    & &  &  -0.05 &   0.17 &   0.45 &   0.31 &   0.49 \\
                 Pb    & &  &   0.25 &   1.92 &   2.31 &   2.16 &   2.46 \\

            \noalign{\smallskip}
            \hline \noalign{\smallskip}
    \end{tabular}
    \caption{Final composition of our models after dilution with the primitive cloud material (see Section~\ref{sec:models}), given in [X/H]. The different initial masses ($M_{ini}$), and masses of the partially mixing zone $\Delta m_{PMZ}$ are shown.}
    \label{tab:yields}
    \end{center}
\end{table*}

\end{appendix}
\end{document}